\documentclass[%
 reprint,
superscriptaddress,
showpacs,preprintnumbers,
nofootinbib,
 amsmath,amssymb,
 bm,
 aps,
]{revtex4-1}
\newcommand{\be}{\begin{eqnarray}}

\newcommand{\ee}{\end{eqnarray}}
\usepackage{graphicx}
\usepackage{amssymb}
\usepackage[utf8]{inputenc} 
\usepackage[T1]{fontenc}    
\usepackage[pdfencoding=auto,psdextra]{hyperref}       
\usepackage{url}            
\usepackage{booktabs}       
\usepackage{amsfonts}       
\usepackage{nicefrac}       
\usepackage{microtype}      
\usepackage{amssymb}
\usepackage{amsmath}
\usepackage{graphicx}
\usepackage[caption=false]{subfig}
\usepackage{changepage}
\usepackage{xcolor}
\usepackage{aas_macros}
\usepackage{multirow}

\begin{document}

\captionsetup[subfigure]{labelformat=empty}

\preprint{APS/123-QED}

\title{Cosmology with Rayleigh Scattering of the Cosmic Microwave Background}

\newcommand{\dampt}{DAMTP, Centre for Mathematical Sciences, Wilberforce Road, Cambridge, UK, CB3 0WA}
\newcommand{\VSI}{Van Swinderen Institute for Particle Physics and Gravity,\\ University of Groningen,
Nijenborgh 4, 9747 AG Groningen, The Netherlands}
\newcommand{\SMU}{Department of Physics,
Southern Methodist University, 3215 Daniel Ave, Dallas, Texas 75275, USA}
\newcommand{\cornell}{Department of Astronomy, Cornell University, Ithaca, New York, USA}

\author{Benjamin Beringue}
\affiliation{\dampt}
\author{P.\ Daniel Meerburg}
\affiliation{\VSI}
\author{Joel Meyers}
\affiliation{\SMU}
\author{Nicholas Battaglia}
\affiliation{\cornell}

\date{\today}
\begin{abstract}
 
The cosmic microwave background (CMB) has been a treasure trove for cosmology. Over the next decade, current and planned CMB experiments are expected to exhaust nearly all primary CMB information. To further constrain cosmological models, there is a great benefit to measuring signals beyond the primary modes. Rayleigh scattering of the CMB is one source of additional cosmological information.  It is caused by the additional scattering of CMB photons by neutral species formed during recombination and exhibits a strong and unique frequency scaling ($\propto \nu^4$). We will show that with sufficient sensitivity across frequency channels, the Rayleigh scattering signal should not only be detectable but can significantly improve constraining power for cosmological parameters, with limited or no additional modifications to planned experiments. We will provide heuristic explanations for why certain cosmological parameters benefit from measurement of the Rayleigh scattering signal, and confirm these intuitions using the Fisher formalism. In particular, observation of Rayleigh scattering allows significant improvements on measurements of $N_{\rm eff}$ and $\sum m_\nu$. 
\end{abstract}

\pacs{Valid PACS appear here}
\maketitle

\section{Introduction}
\label{sec:introduction}
In the current era of precision cosmology, the Cosmic Microwave Background (CMB) has provided the most stringent constraints on the now standard model of cosmology: $\Lambda$CDM. Anisotropies in the temperature of the CMB have been extensively studied over the past decades~\cite{2013ApJS..208...20B,PhysRevD.95.063525,Simard_2018,Akrami:2018vks}. Their angular power spectrum has been measured with increasing accuracy, reaching cosmic variance limits on angular scales larger than $\ell \thicksim 1500$ thanks to the {\it Planck} satellite~\cite{Akrami:2018vks}. CMB polarization (both curl-free $E$- and divergence-free $B$-mode polarization) is another powerful source of information which will be targeted by the next generation of ground based CMB experiments~\cite{Ade_2019,Stacey:2018yqe,Abazajian:2016yjj}. Observationally, two aims of these experiments are to precisely measure the $E$-mode polarization to much smaller scales; and second, to search for $B$ modes of primordial origin~\cite{collaboration2019simons,Abazajian:2019eic}.

With the help of these experiments, we expect nearly all primary CMB temperature and $E$-mode fluctuations over about half of the sky and for multipoles less than a few thousand 
to be mapped in the next decade. Even though more information is carried by modes at even smaller scales, precise measurement of those modes is hampered by the exponential damping of small scale fluctuations and by astrophysical foregrounds. 
In an attempt to continue to develop our cosmological model and extensions thereof, observations
will need to expand beyond the primary CMB fluctuations well-described by linear perturbation theory. Secondary anisotropies in the CMB refer to fluctuations generated beyond linear order in perturbation theory and broadly describe interactions between (primary) CMB photons and cosmic structures. These interactions typically affect the frequency, energy, or direction of propagation of the (primary) CMB photons.

There are various distinguishable ways that cosmic structures can alter the properties of CMB photons~\cite{Aghanim_2008}.
First, photons can interact with gravitational potentials. Two well-known examples are the Integrated Sachs-Wolfe (ISW) Effect and gravitational lensing of the CMB. The ISW effect~\cite{1967ApJ...147...73S} arises when CMB photons travel through a time-varying gravitational potential and thereby acquire a gravitational redshift or blueshift. The gravitational potential will also modify the direction of propagation of the photons. This effect, known as gravitational lensing~\cite{Hu:2000ee,Lewis:2006fu}, deflects primary CMB photons into and out of the line of sight. It changes the statistical properties of the CMB (introducing correlations between modes of different wavenumber). At the level of the power spectrum, the acoustic peaks are smoothed out and large scale power is transferred to small scales.  Another effect on the CMB, called the moving-lens effect, is caused by the motion of gravitational potentials transverse to our line of sight, and should soon be detectable~\cite{1983Natur.302..315B,Cooray_2002,Hotinli:2018yyc}.

Photons can also interact with cosmic structures via scattering off free electrons. Reionization of the Universe by the first stars produces free electrons that are responsible for a suppression of temperature fluctuations on small scales (screening) as well as the generation of $E$-mode polarization on large scales (scattering)~\cite{PhysRevD.79.043003,PhysRevLett.119.021301,Roy:2018gcv}. Both scattering and screening have a linear component, but can also introduce secondaries due to the patchy nature of reionization~\cite{Hu:1999vq,McQuinn:2005ce,Dvorkin:2008tf,Dvorkin:2009ah}. At later time, free electrons can also be encountered in hot gas in galaxy clusters where they give rise to the Sunyaev-Zel'dovich (SZ) effects~\cite{1970Ap&SS...7....3S,1980ARA&A..18..537S,1990PhRvD..41..354B,1995ARA&A..33..541R}. 

Improved sensitivity of current and future CMB experiments has made these secondaries forefront CMB science. Their dual nature, both as an astrophysical nuisance and as a cosmological target, requires exquisite modelling and theoretical understanding. 

In this paper, we will focus on another source of additional cosmological information: Rayleigh scattering of the CMB~\cite{10.1143/ptp/86.5.1021}. Rayleigh scattering generically describes the scattering of photons off neutral species, and its role in the context of the CMB has been studied in various papers~\cite{10.1143/ptp/86.5.1021,Yu_2001,Lewis_2013,PhysRevD.91.083520}. It is usually assumed that after recombination, the Universe becomes transparent to CMB photons. However, a small fraction of these photons can scatter off neutral species formed during recombination leaving a characteristic imprint on the observed temperature and polarization power spectra. Unlike the Thomson scattering-induced primary signal, Rayleigh scattering is strongly frequency dependent. This makes the effect distinguishable from the primary signal. However, in an expanding Universe, the amplitude of the signal is expected to dilute rapidly over time.  The signal scales as $S(\nu) \propto \sigma_R(\nu) \rho(\nu)$, where $\sigma_R(\nu) \propto \nu^4$ (to lowest order) is the Rayleigh scattering cross section for photons of frequency $\nu$, and $\rho(\nu) \propto a^{-3}$ is the density of neutral species; photon frequencies redshift with the expansion of the Universe $\nu\propto a^{-1}$, with $a$ the scale factor. It was shown that at most the Rayleigh signal can reach $3\%$ of the total intensity on the sky~\cite{Yu_2001}. As a result, the Rayleigh signal, while definitely present in the CMB for a standard cosmology, has not yet been detected.  

In previous studies, it was estimated that in principle {\it Planck} could have detected the signal at a few sigma~\cite{Lewis_2013}, while future missions will definitively detect it if enough frequency channels are available. In another paper~\cite{PhysRevD.91.083520} further studies were performed to explore the potential of the signal for cosmology. In particular, as Rayleigh scattering peaks at a nearby but slightly later time than Thomson scattering of the CMB, parameters that change the recombination history, such as the helium fraction $Y_\mathrm{He}$, can benefit from measuring the Rayleigh signal. In addition, the delayed Rayleigh scattering can in principle also provide a better measurement of the expansion history around recombination, allowing for improved measurements of the matter densities in the Universe. 

Together with the improved parameter constraints and the rising costs of extracting further information from the primary CMB, we argue that there is a very strong motivation to detect the Rayleigh scattering spectrum with future CMB experiments. In principle, no special instrumentation is required to extract the signal, and in that sense, an experiment already equipped to measure the primary signal will be suitable to measure the Rayleigh signal as long as multiple frequency channels are available. Among the main obstacles in measuring the Rayleigh signal are galactic and extra-galactic astrophysical foregrounds. Unfortunately, these foregrounds also have a strong scaling with frequency, where both the Cosmic Infrared Background (CIB) and (polarized) dust will be the main sources of confusion. On the other hand, we do in principle know how these foreground components scale with frequency, and by combining multiple frequency channels, it should be possible to isolate the Rayleigh and primary CMB signals. For the purpose of this paper, we will focus on the cosmological benefits of measuring Rayleigh scattering, leaving the practical challenges posed by astrophysical foregrounds to future work.

With multiple CMB experiments planned in the near future that all will aim to map the sky in bands $\sim 50-1000$ GHz~\cite{Ade_2019,Stacey:2018yqe,Abazajian:2016yjj,PICO19}, it is timely to both further explore the detectability and the potential applications of CMB Rayleigh scattering. In the first part of this paper we will review the physics of Rayleigh scattering. We will then present heuristic arguments for the origin of the improvements of parameter constrains that come from measuring Rayleigh scattering of the CMB. 
We will then proceed to forecast both the detectability and parameter constraints for future CMB experiments.

\section{Rayleigh scattering of the Cosmic Microwave Background}
\label{sec:rayleigh}
Rayleigh scattering~\cite{rayleigh_lord_1881_1431155} refers to the frequency-dependent scattering of long-wavelength electromagnetic waves by polarizable particles. The incoming wave excites the internal dipole of the particle which radiates in return, creating an apparent scattering. The strength of scattering scales as $\nu^4$ (to lowest order). This unique frequency scaling is responsible for the sky being blue and sunsets being red~\cite{rayleigh_lord_1899_1431249}. In this section, we will review the cosmological implications of this scattering mechanism around recombination. 

\subsection{Physics of Rayleigh scattering}
\label{subsec:basics_physics}

The process of recombination is accurately described and modelled by well understood physics, and the relevant atomic properties have been extensively studied in laboratory experiments~\cite{Peebles:1968ja,Zeldovich:1969en,Seager:1999bc,Seager:1999km,HyRec11,Chluba16}.  
The most basic model is described by only two quantities: the number density of free electrons in the plasma $n_e$ and the Thomson scattering cross section $\sigma_T$. Before recombination, photons were kept in thermal equilibrium with the plasma through Thomson scattering ensuring that their mean free path remained much smaller than the Hubble horizon.

As the Universe expanded and cooled, the formation of neutral atoms was thermally favored and the number density of free electrons dropped significantly. This made Thomson scattering events less likely and increased the photon mean free path in the plasma. Photons eventually experienced a last scattering event. 
We define the visibility function $g(z) \equiv \dot{\tau}e^{-\tau}$, where $\dot{\tau}$ is the comoving opacity and $\tau$ the optical depth. The visibility, shown in Fig.~\ref{fig:visibility}, should be interpreted as the probability distribution of the last scattering event.

Once the Universe became sufficiently neutral, photons free-streamed and were unlikely to experience scattering until the Universe became reionized as stars and galaxies formed at much lower redshift ($z\lesssim 10$)~\cite{Barkana:2000fd}. However, recombination also produced hydrogen and helium by which photons can also be scattered through \emph{Rayleigh scattering}. This process is governed by the Rayleigh scattering cross section that can be written as~\cite{PhysRevD.91.083520,Lewis_2013}:
\begin{equation}
        \sigma_R(\nu) = \sigma_T \left[\sum_{j = 2}^{\infty}f_{1j}\frac{\nu^2}{\nu_{1j}^2 - \nu^2}\right]^2, 
    \label{eq:rayleigh_cross1}
\end{equation}
where $\nu_{1j}$ and $f_{1j}$ are the Lyman series frequencies and oscillator strengths and $\sigma_T$ is the Thomson scattering cross section. Note that the cross section for Rayleigh scattering depends on the frequency of the photons while Thomson scattering does not.

Around recombination, it is a good approximation to treat typical CMB photons as having frequency much smaller than any of the Lyman series transitions. By defining $\nu_{\rm eff} = \sqrt{8/9}R_\infty c \approx 3.102\times 10 ^{6}~{\rm GHz}  \approx 12.83~\mathrm{eV}$, with $R_\infty$ the Rydberg constant, we can expand Eq.~\eqref{eq:rayleigh_cross1} as follows:    
\begin{equation}
       \sigma_R(\nu) \approx \sigma_T\left[\left(\frac{\nu}{\nu_{\rm eff}}\right)^ 4 + \alpha\left(\frac{\nu}{\nu_{\rm eff}}\right)^ 6 + \beta \left(\frac{\nu}{\nu_{\rm eff}}\right)^ 8 + ... \right], 
       \label{eq:rayleigh_cross2}
\end{equation}
where $\alpha \approx 2.626$ and $\beta \approx 5.502$ (exact values can be found in~\cite{Lewis_2013}). We recover the leading $\nu^4$ scaling (which is familiar from scattering of solar photons by our atmosphere) with higher-order contributions becoming relevant at higher frequencies. After recombination, densities of the neutral species will evolve as $(1+z)^3$ and the frequency of CMB photons as $(1+z)$, making the probability of Rayleigh scattering $\propto (1+z)^7$ (at leading order). Consequently, Rayleigh scattering events remain localised around recombination.

Notice that on average, Rayleigh scattering is as likely to scatter photons into our line of sight as it is to scatter photons out of our line of sight.  As a result, there is no monopolar distortion of the CMB frequency spectrum due to Rayleigh scattering, just as Thomson scattering by free electrons present after reionization does not change the mean temperature of the CMB.

Rayleigh scattering induces two major changes to the recombination history. First, it increases the overall coupling between baryons and photons, and second, it introduces a frequency dependence. We will briefly review the consequences of these changes in the next section.

\subsection{Effects of Rayleigh scattering on the recombination history.}
\label{subsec:effects}
\begin{figure}
{\centering
  \includegraphics[width = \columnwidth]{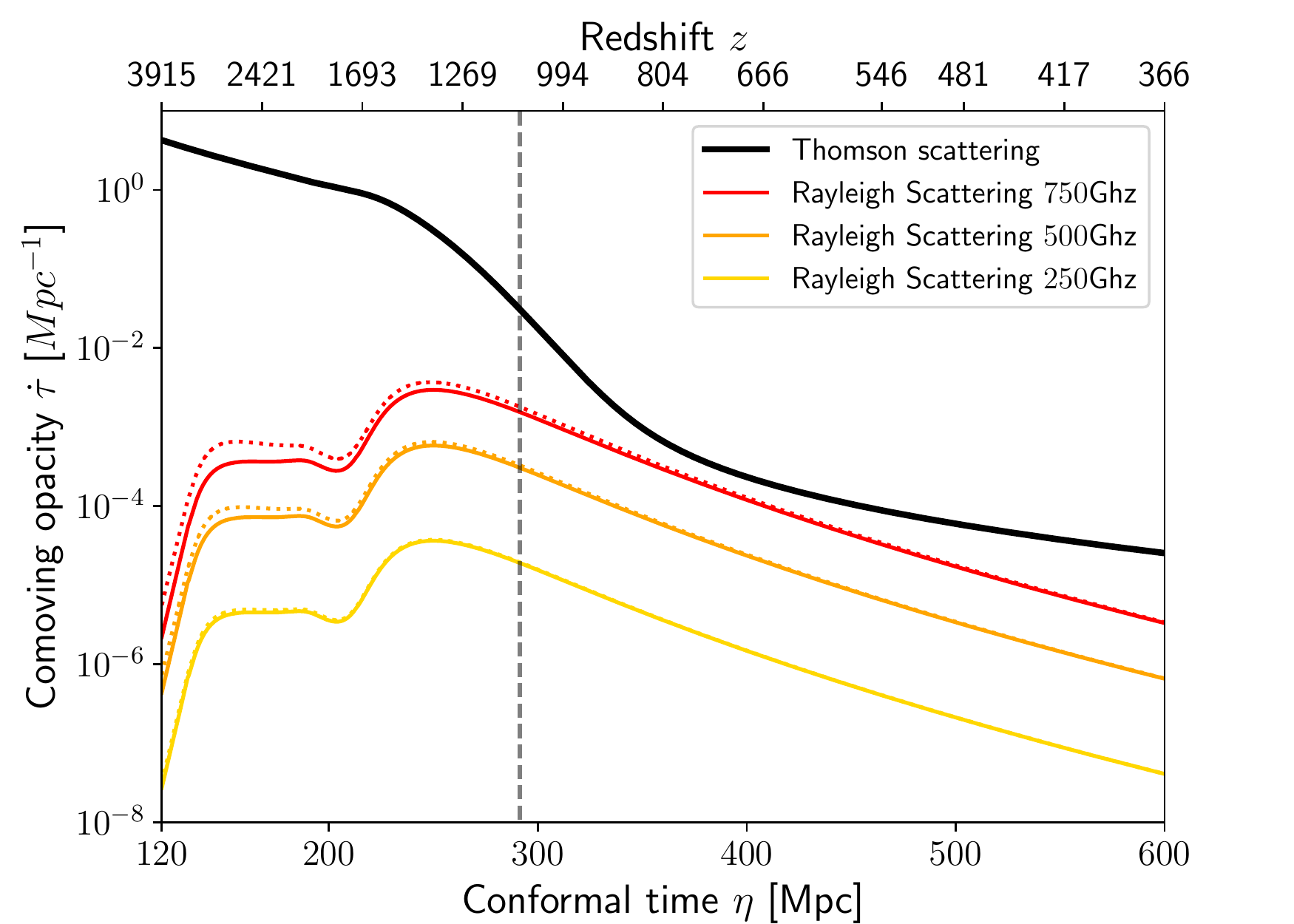}
}
  \caption{Contributions to the comoving opacity for Thomson scattering (black) and Rayleigh scattering for different frequencies at the leading $\nu^4$ order. Dotted lines represent contributions including both $\nu^4$ and $\nu^6$ orders. The vertical dashed line shows the last scattering surface.}
   \label{fig:opacity}
\end{figure}

\paragraph{Increased comoving opacity}
Including Rayleigh scattering in the CMB photons Boltzmann equation increases the comoving opacity in a frequency-dependent way, i.e. 
\begin{align}\label{eq:opacity}
    \dot{\tau} &= an_e\sigma_T  \nonumber \\
    \rightarrow \dot{\tau}(\nu) &= an_e\sigma_T + a(n_H + R_\mathrm{He}n_\mathrm{He})\sigma_R(\nu),
\end{align}
 where $R_\mathrm{He} \approx 0.1$ accounts for Rayleigh scattering by Helium atoms being less efficient than Hydrogen~\cite{Lewis_2013,PhysRevD.91.083520}. Note that heavier elements as well as ionized species also scatter photons but their contribution to Rayleigh scattering is suppressed by their low abundance. Fig.~\ref{fig:opacity} shows the evolution of the comoving opacity $\dot{\tau}$ for both Thomson and Rayleigh scattering as a function of redshift. The Rayleigh scattering opacity increases around recombination ($z\thicksim 1100$) as neutral hydrogen becomes more abundant and decreases after recombination ($ \propto (1+z)^{7}$) when the photon frequency redshifts and neutral hydrogen becomes more dilute. \\

\begin{figure*}[!ht]
\centering
\subfloat[]{
    \includegraphics[width=0.475\linewidth]{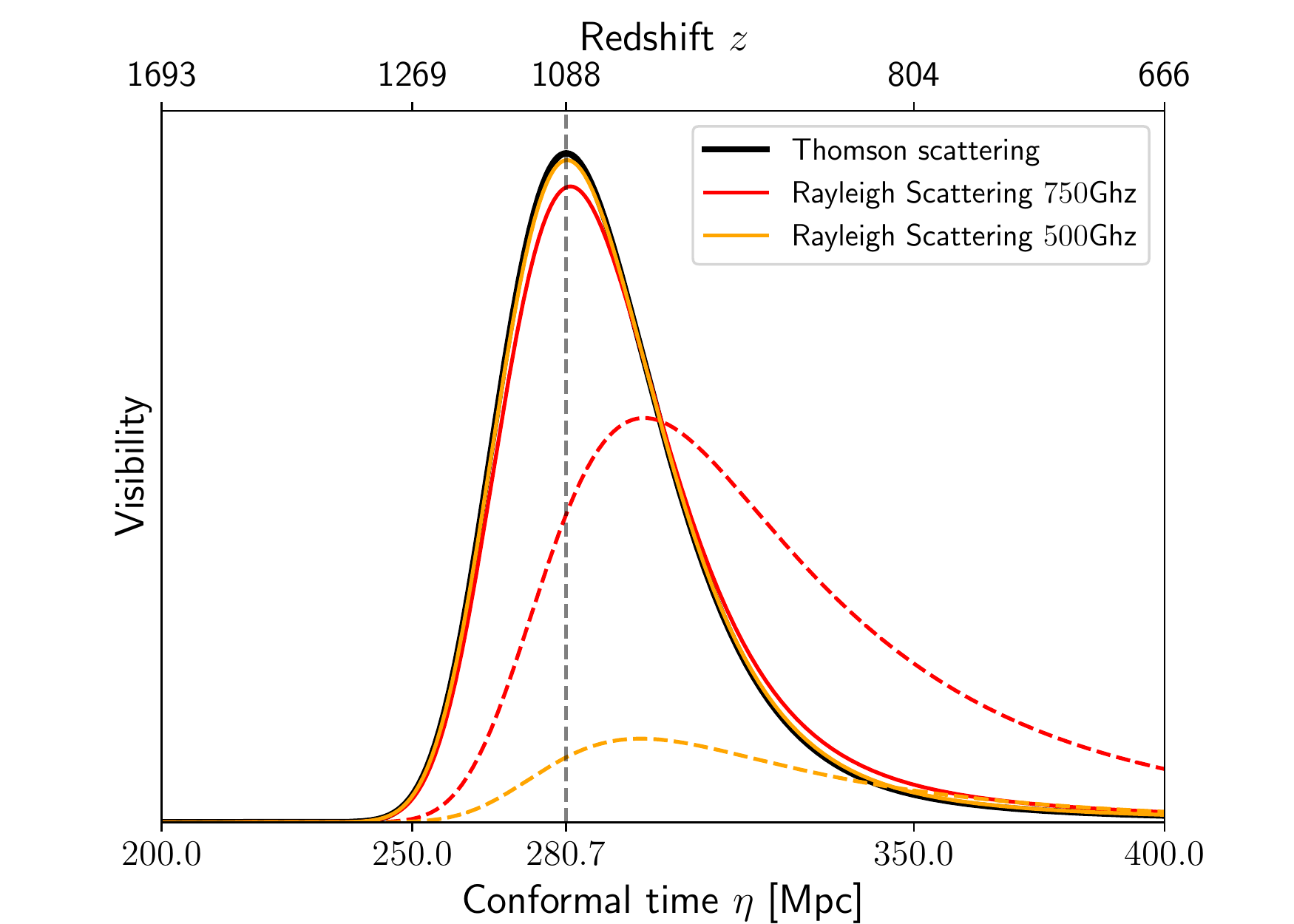}
}
\hfill
\subfloat[]{
    \includegraphics[width=0.475\linewidth]{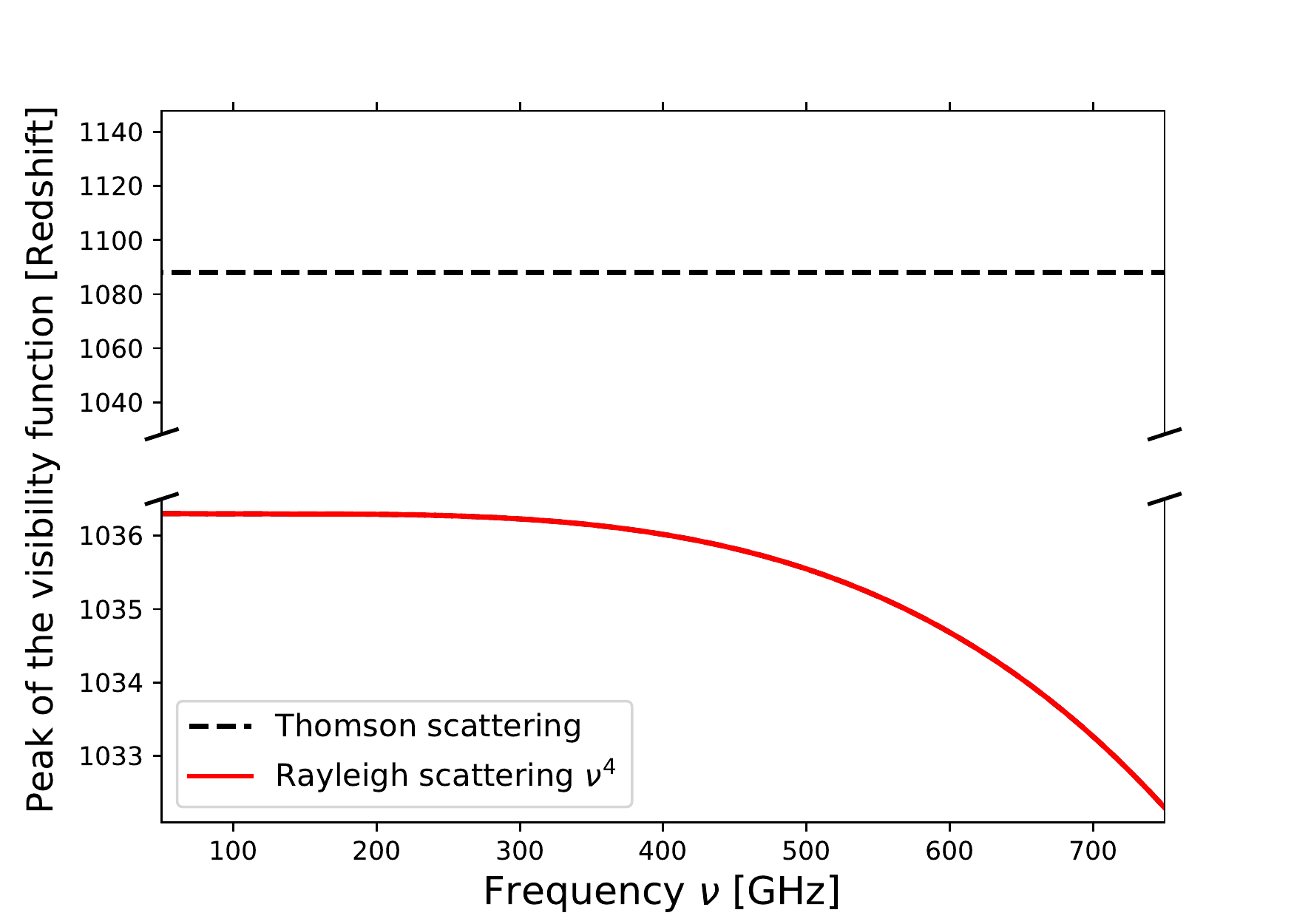}
}
\caption{\textit{Left:} Changes to the visibility function $g(z) = \dot{\tau}e^{-\tau}$ due to Rayleigh scattering for two frequencies. Rayleigh scattering tends to move the peak of the visibility function (i.e. the surface of last scattering) towards lower redshift. Dashed lines correspond to the (normalized) contributions of Rayleigh scattering to the total visibility (solid). \textit{Right:} Location of the peak of the visibility function for the Thomson (dashed black) and Rayleigh (solid red) scattering terms.}
\label{fig:visibility}
\end{figure*}

\paragraph{Shift of the visibility function} The visibility function is defined as:
\begin{equation}
    g\left(z\right) = \dot{\tau}e^{-\tau}.
    \label{eq:visibility}
\end{equation}
Due to Rayleigh scattering, the visibility function becomes frequency dependent. More importantly, since the total coupling between baryons and photons is increased, the last scattering event (irrespective of it being Thomson scattering on a free electron or Rayleigh scattering on a neutral species) will be shifted towards later times. This leads to a (frequency-dependent) shift of the visibility function towards later times (lower redshift), as illustrated in Fig.~\ref{fig:visibility}. \\

\paragraph{Increase of diffusion damping} 
The amplitude of diffusion damping~\cite{Hu97} is directly controlled by the photon mean free path in the plasma. The shorter the mean free path, the smaller the effect. As described earlier, the photon mean free path is shortened by Rayleigh scattering which reduces the diffusion length and consequently the amplitude of diffusion  damping.  While this holds at low frequency, at higher frequency, Rayleigh scattering also shifts the visibility function towards later times where diffusion  Damping is stronger (the mean free path globally increases with time after recombination) leading to an overall increase in the amplitude of diffusion Damping. \\

\paragraph{Frequency-dependent sound horizon at last scattering}
As mentioned above, the last scattering surface becomes frequency dependent due to Rayleigh scattering. Therefore the size of the sound horizon at last scattering also becomes frequency dependent, being larger at higher frequencies. The size of the sound horizon dictates the location of the acoustic peaks in both the matter and the CMB power spectra, a larger sound horizon shifts the acoustic peaks towards larger scales. \\

\begin{figure*}[!t]
\centering
\subfloat[]{
\includegraphics[width=0.475\linewidth]{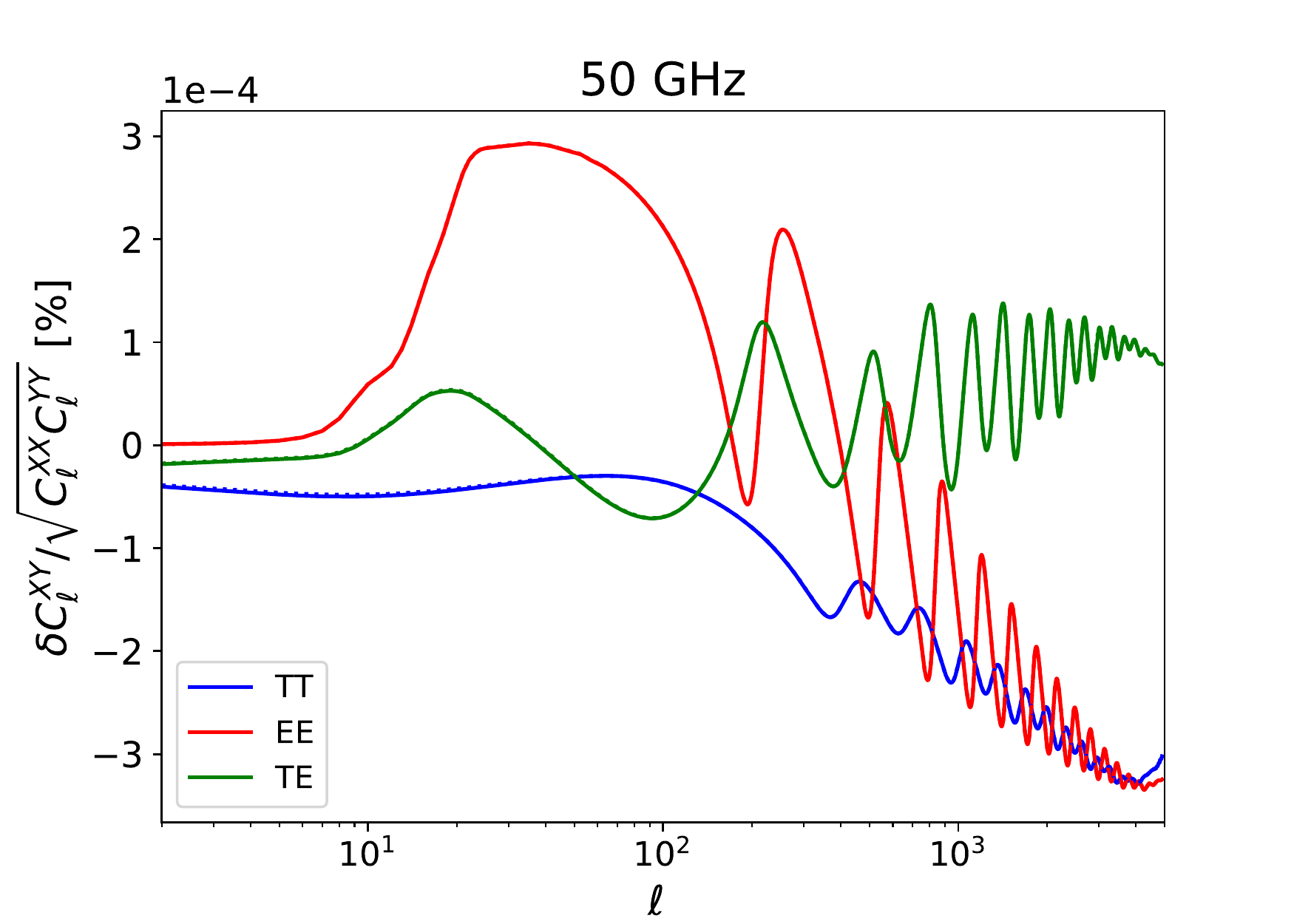}
}
\hfill
\subfloat[]{
\includegraphics[width=0.475\linewidth]{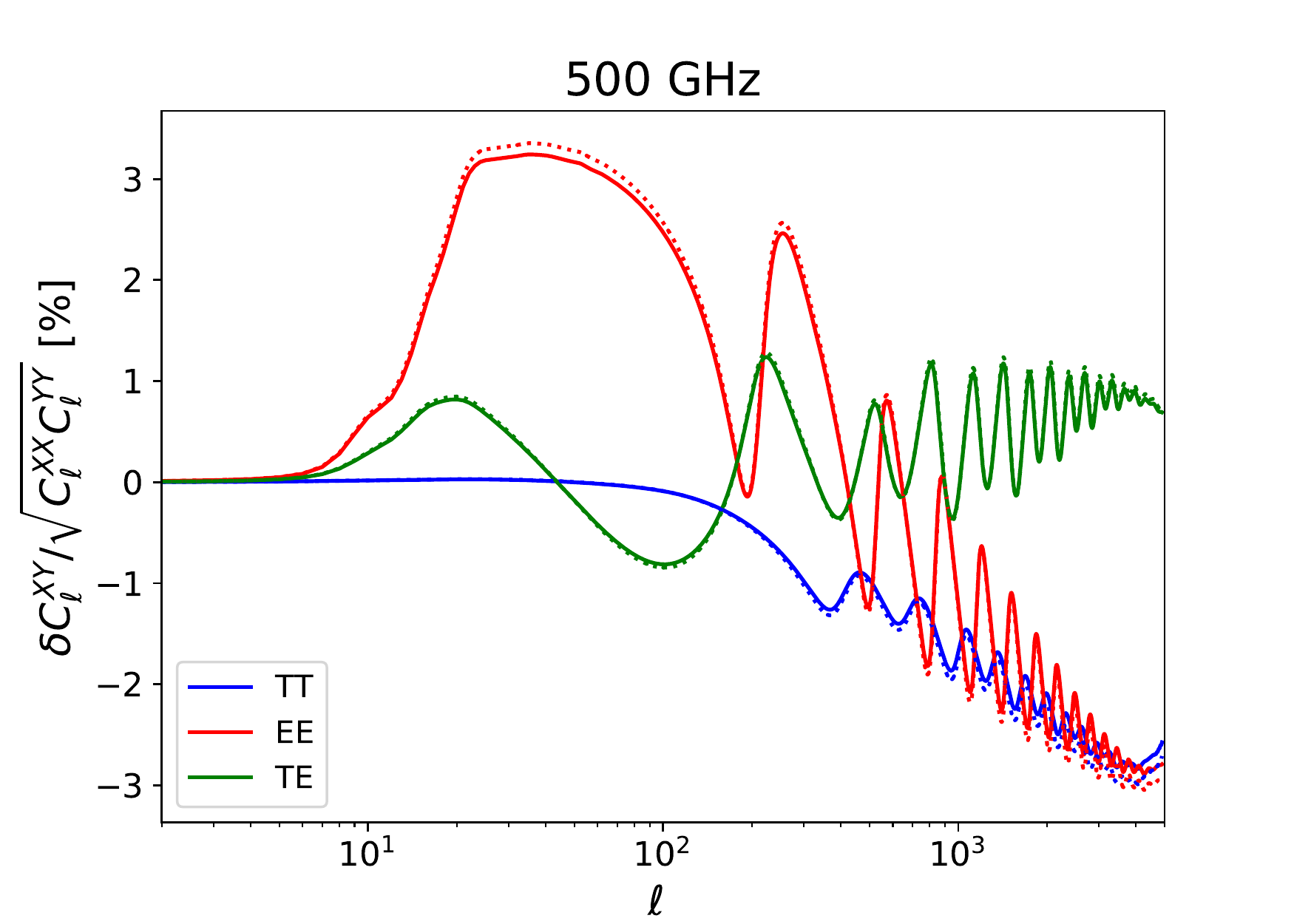}
}
    \caption{Fractional difference $\left(\delta C_\ell^{XY} / \sqrt{C_\ell^{XX}C_\ell^{YY}}\right)$ induced by Rayleigh spectra in lensed $TT$, $TE$, and $EE$ CMB spectra. On small scales, both $TT$ and $EE$ experience additional damping by Rayleigh scattering. On large scales, the auto spectrum of $E$ mode polarization is boosted. Oscillations are due to the shift in the location of the acoustic peaks induced by Rayleigh scattering. Dashed lines include the effects of the $\nu^6$ term, which is negligible for the frequencies considered.}
    \label{fig:fractional_diff}
\end{figure*}

\paragraph{Frequency-dependent amplitude of polarization signal}
CMB radiation is weakly linearly polarized~\cite{PhysRevD.51.364,Kosowsky_1996,Zaldarriaga_1997,Kamionkowski_1997,Hu_1997}. CMB polarization can be projected onto two orthogonal modes, the familiar curl-free $E$ and gradient-free $B$ polarization modes. While primary $B$ modes are not generated at linear order by primordial density perturbations, $E$ modes are sourced by scalar fluctuations as scattering of quadrupole temperature anisotropies present around recombination produce linear polarization. The shift of the visibility function induced by Rayleigh scattering increases the amplitude of the local temperature quadrupole around the time of recombination, leading to a boost in $E$ modes on large scales as presented in Fig.~\ref{fig:fractional_diff}. \\

To summarize, Rayleigh scattering is responsible for: 
\begin{itemize}
    \item A damping of small scales anisotropies both in temperature and $E$-mode polarization. This is due to an increase of diffusion  damping.
    \item On large angular scales, Rayleigh scattering primarily affects the  $E$-mode polarization signal. By shifting the last scattering surface towards lower redshifts, where the local quadrupole is larger, Rayleigh scattering boosts the large scale $E$-mode signal. 
    \item Rayleigh scattering introduces frequency dependence in the size of the sound horizon, leading to a shift in the location of the acoustic peaks, both in temperature and $E$-mode polarization spectra.
\end{itemize} 

Fig.~\ref{fig:fractional_diff} shows the fractional differences introduced by Rayleigh scattering on the $TT$, $TE$, and $EE$ CMB spectra.\footnote{All the CMB spectra used in this article have been computed using the Rayleigh branch of CAMB, \url{http://camb.info}~\cite{Lewis99}.} 

\subsection{Modeling Rayleigh scattering distortions}
\label{subsec:modeling}
In the range of frequencies of interest for cosmological analysis ($\nu \in [20,800]~\text{GHz}$), Refs.~\cite{PhysRevD.91.083520,Lewis_2013} showed that the distortions induced by Rayleigh scattering can be accurately captured by an additional random variable for each term in Eq.~\eqref{eq:rayleigh_cross2}. In this section, we present the model we will use throughout this article and leave the discussion on additional choices to Appendix~\ref{app:eigenvalues}.  

Power spectra are defined in terms of the angular multipoles of the signal of interest: 
\begin{equation}
    C_\ell^{XY} = \langle a_{\ell m} ^ X  a_{\ell m} ^ Y \rangle, 
\end{equation}
where $X$ and $Y$ can be $T$ or $E$ (we don't include B-modes in this analysis since they are either sourced by gravitational lensing or tensor fluctuations which are respectively not impacted by Rayleigh scattering and not yet detected). When we include Rayleigh scattering, these quantities become frequency dependent. Our first approach consists in adding distortions to the primary signals at each frequency~\cite{Lewis_2013}, i.e., 
\begin{equation}
    a_{\ell m}^X \left( \nu \right) \thicksim a_{\ell m}^X + \left( \frac{\nu}{\nu_0}\right)^4 \Delta a_{\ell m}^{X,4} + \left( \frac{\nu}{\nu_0}\right)^6 \Delta a_{\ell m}^{X,6} + \ldots \, ,
    \label{eq:signal_rs}
\end{equation}
where $\nu_0$ is a reference frequency and $\Delta a_{\ell m}^{X,4}$ and $\Delta a_{\ell m}^{X,6}$ account for the distortions introduced by Rayleigh scattering to the primary signal $a_{\ell m}^X$. At very high frequencies ($\nu  \gtrsim 800$GHz), higher order terms will become important and so will the frequency dependence of the optical depth around recombination. This will break the assumption that Rayleigh scattering acts as a linear perturbation as in Eq.~\eqref{eq:signal_rs}. However at these frequencies there are very few CMB photons due to the blackbody nature of the CMB radiation. Furthermore, bright galactic and extra-galactic foregrounds will likely restrict use of such high frequencies to measure the Rayleigh signal, and we do not consider them further in this work. 

\begin{figure*}[!ht]
\subfloat[]{
    \includegraphics[width=0.475\linewidth]{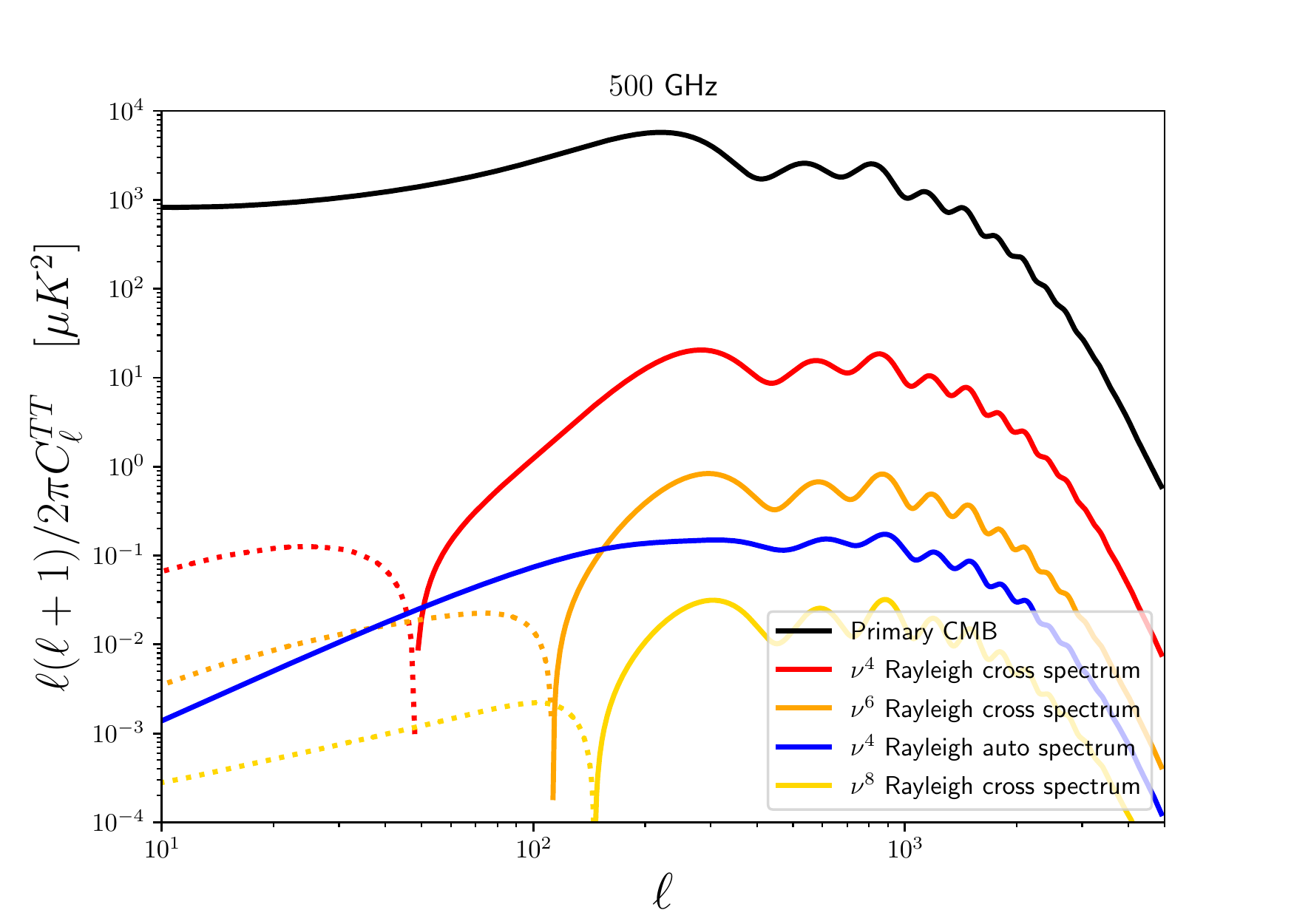}
}
\hfill
\subfloat[]{
    \includegraphics[width=0.475\linewidth]{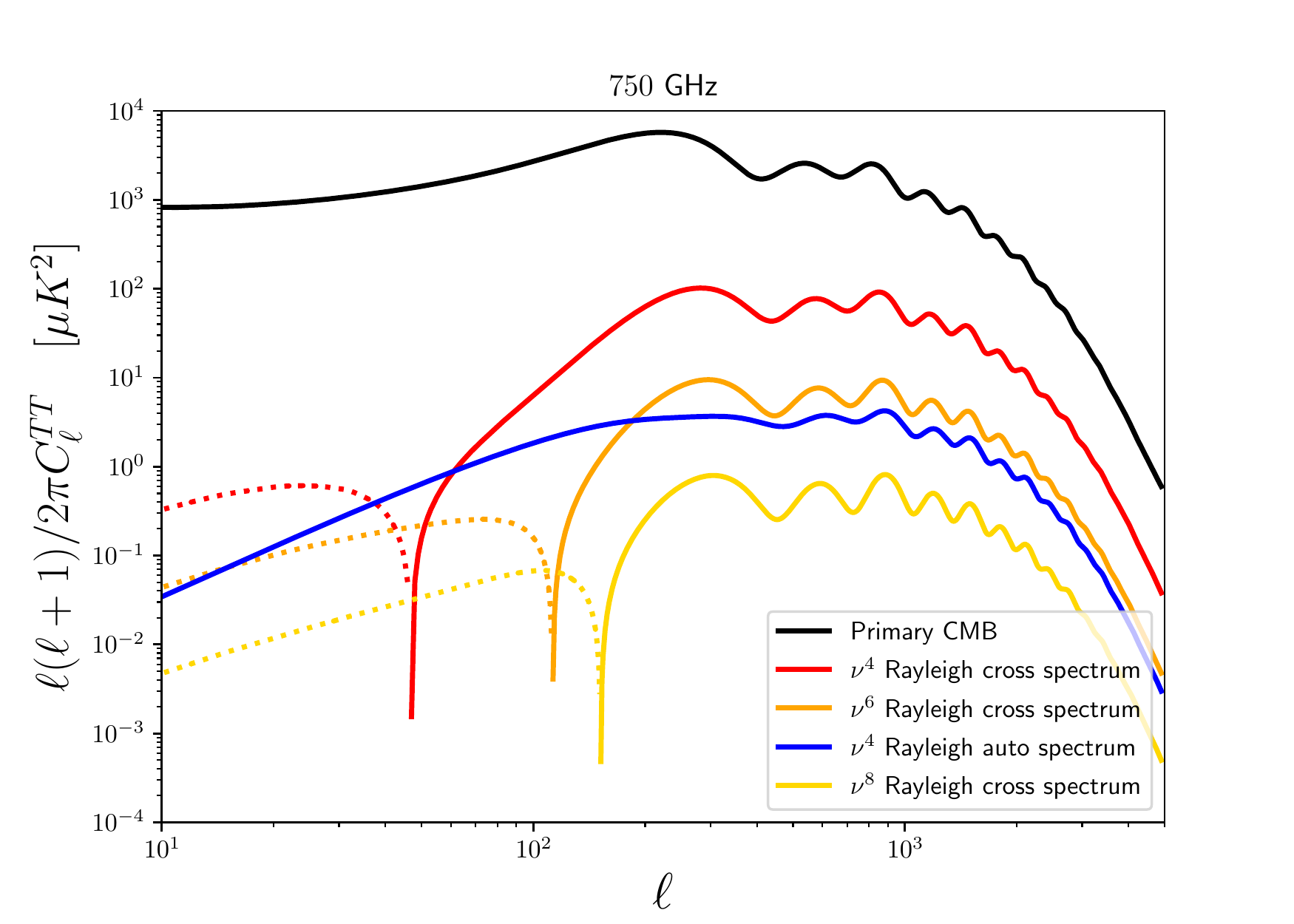}
}
\caption{Various terms in the Rayleigh scattering spectra expansion (Eq.~\eqref{eq:rs_expansion}) for two frequencies: $500$~GHz (left), $750$~GHz (right). Shown are $TT$ spectra: primary CMB (black), $\nu^4$ Rayleigh cross spectrum (red),  $\nu^6$ Rayleigh cross-spectrum (orange),  $\nu^4$ Rayleigh auto spectrum (blue), and  $\nu^8$ Rayleigh cross-spectrum (yellow). Solid lines in the cross-spectra show negative correlation and dotted lines positive correlation. 
}
\label{fig:rayleigh_terms}
\end{figure*}

For the remainder of this discussion we will therefore assume the effect of Rayleigh scattering on the CMB power spectra can be accurately modeled by Eq.~\eqref{eq:signal_rs}. This leads to power spectra that can be written as~\cite{Lewis_2013} 
\begin{align}
        C_{\ell}^{XY}\left(\nu_1,\nu_2\right) &= \left< a_{\ell m} ^ X \left(\nu_1\right) a_{\ell m} ^ Y \left(\nu_2\right)\right> \nonumber\\
        &= C_{\ell}^{XY} + \left(\frac{1}{\nu_0}\right)^4\left[\nu_1^4 C_{\ell}^{X\Delta Y_4} + \nu_2^4 C_{\ell}^{\Delta X_4 Y} \right] \nonumber\\
        &\qquad + \left(\frac{1}{\nu_0}\right)^6\left[\nu_1^6 C_{\ell}^{X\Delta Y_6} + \nu_2^6 C_{\ell}^{\Delta X_6 Y} \right] \nonumber\\
        &\qquad + \left(\frac{1}{\nu_0}\right)^8\left[\nu_1^8 C_{\ell}^{X\Delta Y_8} + \nu_2^8 C_{\ell}^{\Delta X_8 Y} \right] \nonumber\\
        &\qquad + \left(\frac{\nu_1\nu_2}{\nu_0^2}\right)^4\left[C_{\ell}^{\Delta X_4 \Delta Y_4}\right] + \ldots
\label{eq:rs_expansion}
\end{align}
The first term in this expansion is the primary CMB, sourced by Thomson scattering. the following three are cross-correlation spectra between the primary CMB and the Rayleigh scattering signal. The last term gives the Rayleigh scattering auto-spectrum from the first term in Eq.~\eqref{eq:rayleigh_cross2}. We display each of these contributions in Fig.~\ref{fig:rayleigh_terms}. 

Fig.~\ref{fig:rayleigh_terms} also shows that the Rayleigh signal lies significantly below the primary signal, indicating that it will be challenging to detect. The cross spectra will initially be the most promising way to observe Rayleigh scattering. In particular, $TE$ cross-spectra would further benefit from limited foreground contamination. We leave a more detailed discussion of detectability to Section \ref{sec:forecasts}. \\

\begin{figure}[ht]
    \centering
    \includegraphics[width = .5\textwidth]{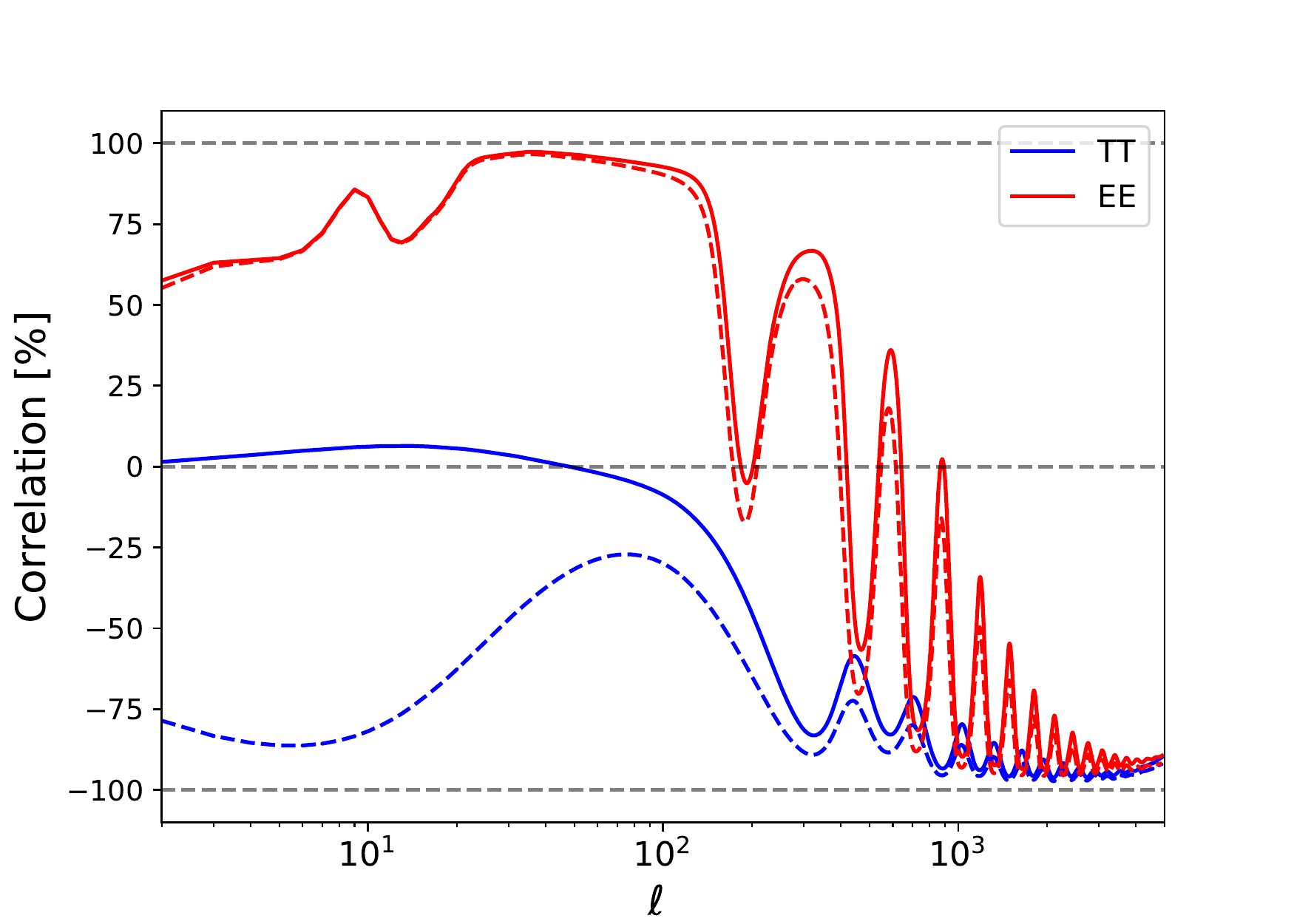}
    \caption{Correlation coefficient of the primary-Rayleigh cross spectrum (defined as $r\equiv C_{\ell}^{\Delta X_4 X} /  \sqrt{C_{\ell}^{XX}C_{\ell}^{\Delta X_4 \Delta X_4}}$) for $X = T$ (blue) and $X=E$ (red) and for two frequencies, $50$~GHz (dashed) and $500$~GHz (solid). In polarization, large scales are boosted because of the larger local temperature quadrupole near the peak of the Rayleigh scattering visibility function, hence the positive correlation. Large scales in temperatures are partly (anti-) correlated at lower frequencies and uncorrelated at higher frequencies where the shift in the visibility function becomes substantial. On small scales, Rayleigh scattering distortions are anti-correlated with the primary CMB.}
    \label{fig:correlation}
\end{figure}

\section{Rayleigh scattering and Cosmological Parameters}
\label{sec:cosmo_parameters}

The frequency dependence of Rayleigh scattering results in different CMB power spectra at each observed frequency.  These changes of the spectra with frequency provide some additional information which can be used to constrain cosmological parameters.  In this section, we will describe which parameters are expected to be more tightly constrained by observing the effects of Rayleigh scattering on the CMB, and we will provide a heuristic description of where the additional constraining power comes from.  In Sec.~\ref{sec:forecasts}, we will provide quantitative statements about how much improvement can be derived from observation of Rayleigh scattering with future CMB experiments.

One possibility provided by observation of Rayleigh scattering is access to a set of independent primordial fluctuations, but making use of these new modes would require a measurement of the uncorrelated Rayleigh scattering component, a task which is extremely difficult to achieve especially in the presence of foregrounds~\cite{Lewis_2013}.  There is still much to be gained by observation of the correlated component of Rayleigh scattering which was sourced by the same primordial fluctuations as the primary CMB fluctuations.  We will focus here on the benefits provided by measuring the correlated component, since that is much more tractable in the presence of foregrounds and is a goal within reach of near-future experiments.

It has previously been shown that measurement of the effects of Rayleigh scattering on the CMB can significantly improve the measurement of the primordial helium abundance~\cite{PhysRevD.91.083520}.  This is due to the fact that at the relevant frequencies, the Rayleigh scattering cross section with helium is much lower than that of hydrogen (see Eqs.~\eqref{eq:rayleigh_cross1} and \eqref{eq:opacity}).  As a result, the amplitude of the Rayleigh scattering effects on the CMB scale inversely with the primordial helium abundance.  This scaling would allow one to use observations of CMB Rayleigh scattering to constrain $Y_\mathrm{He}$ even if there were no other effects on the spectra that resulted from changing $Y_\mathrm{He}$.  In fact, changes to the primordial helium abundance alter the damping scale of the CMB power spectra~\cite{Zaldarriaga:1995gi,Bashinsky:2003tk,Hou:2011ec}, and as we will discuss below, this provides another way in which measurements of Rayleigh scattering can better constrain the primordial helium abundance. Furthermore, dark matter which efficiently scatters with baryons in the early Universe (such as milli-charged dark matter) can mimic some of the effects of primordial helium in the CMB~\cite{dePutter:2018xte}, but such dark matter would not be expected to exhibit Rayleigh scattering and could therefore be distinguished through the frequency dependence of CMB power spectra.

We can gain another handle on cosmological parameters from Rayleigh scattering due to the fact that the peak of the visibility function for Rayleigh scattering occurs at a lower (frequency-dependent) redshift than that for Thomson scattering (see Fig.~\ref{fig:visibility} and Sec.~\ref{sec:rayleigh}).  As a result, any physics which causes a feature at a particular length scale will result in a feature at a different angular scale in the CMB for Thomson scattering than for Rayleigh scattering. One example is provided by the positions of the acoustic peaks in the CMB spectra.  The comoving size of the sound horizon at the redshift corresponding to the peak of the Thomson scattering visibility function is $r_\mathrm{s}^\star$.  The spacing between acoustic peaks in the Thomson scattering spectrum is therefore expected to be $\Delta \ell \simeq \pi D_A^\star/r_\mathrm{s}^\star\equiv\pi/\theta_\mathrm{s}^\star$ where $D_A^\star$ is the angular diameter distance to the peak of the visibility function due to Thomson scattering~\cite{Hu:2001bc}.  The peak positions will be different for the Rayleigh scattered photons, for which $\Delta \ell^R(\nu) \simeq \pi D_A^{\mathrm{R}\star}(\nu)/r_\mathrm{s}^{\mathrm{R}\star}(\nu)$ where the superscript R refers to the Rayleigh scattered component.  Both the size of the comoving sound horizon and the angular diameter distance to a particular redshift depend on the cosmological history
\begin{equation}\label{eq:sound_horizon_and_DA}
    r_\mathrm{s}(z) = \int_z^\infty \frac{dz}{H(z)}c_\mathrm{s}(z) \, , \qquad D_A(z)=\int_0^z \frac{dz}{H(z)} \, ,
\end{equation}
where $c_\mathrm{s}(z)$ is the sound speed at redshift $z$.  Measurement of both the primary CMB fluctuations and (cross-correlation with) the Rayleigh scattered component therefore provides additional information about the cosmological parameters which affect the expansion history and sound speed.

As another example, one of the ways in which we can infer the density of non-relativistic matter in the early Universe is to measure the difference in amplitude of fluctuations which entered the horizon before and after matter-radiation equality~\cite{Seljak:1994yz,Hu:1994uz,Hu:1995en,Hu:2001bc}.  The angular multipole at which this transition appears in the primary CMB spectra is $\ell_\mathrm{eq}\sim k_\mathrm{eq}D_A^\star$ where $k_\mathrm{eq}$ is wavenumber of a cosmological fluctuation which enters the sound horizon at matter-radiation equality.  For the Rayleigh scattered spectra, this transition instead occurs at $\ell_\mathrm{eq}^\mathrm{R}(\nu)\sim k_\mathrm{eq}D_A^{\mathrm{R}\star}(\nu)$.  Identifying both $\ell_\mathrm{eq}$ and $\ell_\mathrm{eq}^\mathrm{R}(\nu)$ can therefore better constrain the parameters which affect the matter-radiation equality scale (like $\Omega_c h^2$ and $N_\mathrm{eff}$) and help to break degeneracies that are present with only primary CMB measurements.  This is particularly beneficial for constraints on the sum of neutrino masses $\sum m_\nu$, since the additional constraining power on $\Omega_c h^2$ around the time of recombination can nearly obviate the need for low-redshift BAO measurements to infer the non-relativistic matter density at late times that is usually required to make a good neutrino mass measurement with CMB data~\cite{Abazajian:2016yjj,PICO19,Dvorkin:2019jgs}; see Table~\ref{tab:extensions_forecasts}.  This allows for the possibility that a high significance detection of the minimal sum of neutrino masses ($\sum m_\nu \simeq 60$~meV) could be made with the data from a single CMB experiment, without the need for external data, if it is capable of utilizing Rayleigh scattering information.

The diffusion damping length of cosmological fluctuations provides another physical scale whose angular size can be measured in the CMB and that will differ for pure Thomson scattering and the Rayleigh scattered component.  The damping length is determined by the free electron fraction, expansion history, and baryon fraction prior to recombination~\cite{Hu:1996vq,Hou:2011ec}.  Since the energy density of relativistic species determines the expansion rate at early times, the damping tail is especially sensitive to the parameter $N_\mathrm{eff}$, which describes the density of light relics.  The positions of acoustic peaks on small angular scales are also impacted by a phase shift imparted by fluctuations in the density of free-streaming light relics~\cite{Bashinsky:2003tk,Baumann:2015rya}, which in principle gives an extra handle on $N_\mathrm{eff}$ that can provide improved constraints when measuring Rayleigh scattering. As shown in Sec.~\ref{sec:forecasts} and Table~\ref{tab:extensions_forecasts}, measurements of Rayleigh scattering can provide modest improvements to constraints on $N_\mathrm{eff}$. Even small improvements in the error on $N_\mathrm{eff}$ are extremely valuable, since the energy scale of the new physics that can be probed with such measurements is a very non-linear function of $\sigma(N_\mathrm{eff}$)~\cite{Green:2019glg}. The exponential suppression of small scale CMB fluctuations means that it will be very challenging to significantly improve constraints on cosmological parameters like $N_\mathrm{eff}$ that are best measured with the damping tail by simply adding more detectors to CMB telescopes after the upcoming generation of experiments~\cite{Abazajian:2016yjj,PICO19}. Measurements of Rayleigh scattering thereby provide an important new avenue by which to pursue improvements on measurements of the light relic density.

\section{Forecasts for future generation of experiments}
\label{sec:forecasts}
\subsection{Detectability}
In order to assess the detectability of Rayleigh scattering by future experiments, we will present Fisher forecasts using realistic noise levels of several proposed and planned CMB experiments. We will ignore astrophysical foregrounds and instrument systematics for now, but we do include the effect of the Earth's atmosphere for ground-based experiments. In a forthcoming publication, we will explore component separation techniques to deal with astrophysical foregrounds and systematics mitigation to extract the Rayleigh scattering signal. In this section, we will also neglect $\nu^6$ and higher order corrections in Eq.~\eqref{eq:signal_rs}, which will have negligible effects on the forecasts. 

\subsubsection{Fisher matrix formalism}
\label{subsubsec:fisher_formalism}
Throughout this paper, all the forecasts will be carried out using the Fisher matrix formalism~\cite{PhysRevD.52.4307}. This formalism assumes a Gaussian likelihood and although it may not represent the true likelihood, it provides a fast and accurate method for computing how well future experiments can constrain parameters (see e.g.~\cite{Wu14} for an example of an implementation). Assuming a likelihood $\mathcal{L}(\mathbf{\theta}|\mathbf{d})$ where $\mathbf{\theta}$ is the vector of (cosmological) parameters, $\mathbf{d}$ the data vector, and the theoretical covariance is $\mathbf{C}(\mathbf{\theta})$, the likelihood reads 
\begin{equation}
    \mathcal{L}(\mathbf{\theta}|\mathbf{d}) \propto \frac{1}{\sqrt{\det(\mathbf{C}(\mathbf{\theta}))}} \exp \left( -\frac{1}{2}\mathbf{d}^\dagger \left(\mathbf{C}(\mathbf{\theta})\right)^{-1}\mathbf{d}\right).
    \label{eq:likelihood}
\end{equation}
The fisher matrix is defined as~\cite{PhysRevD.52.4307}:
\begin{equation}
    F_{ij} \equiv - \left \langle \left.\frac{\partial^2\log \mathcal{L}}{\partial\theta_i\theta_j}\right|_{\theta = \theta_0} \right \rangle.
    \label{eq:fisher_def}
\end{equation}
Combining Eqs.~\eqref{eq:likelihood} and \eqref{eq:fisher_def}, we can write the Fisher matrix for a CMB experiment as 
\begin{equation}
    F_{ij,\ell} = \frac{1}{2}\mathrm{Tr} \left[\left(\mathbf{C}_\ell\right)^{-1}\frac{\partial \mathbf{C}_\ell}{\partial\theta_i}\left(\mathbf{C}_\ell\right)^{-1}\frac{\partial \mathbf{C}_\ell}{\partial\theta_j}\right].
    \label{eq:fisher_CMB}
\end{equation}
Since each value of $\ell$ corresponds to $f_{\rm sky}(2\ell +1)$ independent modes (where $f_{\rm sky}$ is the sky coverage of the experiment), the total fisher information is given by
\begin{equation}
    F_{ij} = \sum_\ell f_{\rm sky}(2\ell+1)F_{ij,\ell}. 
    \label{eq:fisher_tot}
\end{equation}

We will take the $\theta_i$ to be amplitudes of the spectra when assessing detectability, and when forecasting parameter constraints, the $\theta_i$ will refer to the cosmological parameters. 

\subsubsection{Detecting cross-correlation}

First, we will forecast the detectability of the cross-correlations between the primary and the Rayleigh signal using the frequency covariance matrix~\cite{Lewis_2013}. As presented in Fig.~\ref{fig:rayleigh_terms}, the most likely avenue to detect the faint Rayleigh scattering signal is through its cross-correlation with the primary CMB. In practice, this would be achieved by looking for a frequency-dependent correlation between a low frequency map (with negligible contribution from Rayleigh scattering) and a foreground-cleaned high frequency map. We write 
\begin{equation}
    \mathbf{C}_\ell^\mathrm{\nu\nu} = 
    \begin{pmatrix}
    \mathbf{C}_\ell^{TT} & \mathbf{C}_\ell^{TE} \\[6pt]
    \mathbf{C}_\ell^{ET} & \mathbf{C}_\ell^{EE}
    \end{pmatrix},
\end{equation}
where each $\mathbf{C}_\ell^{XY}$ are $N_\nu\times N_\nu$ frequency-covariance matrices, where $N_\nu$ is the number of frequencies that are measured.  Each entry of these matrices  are defined in Eq.~\eqref{eq:rs_expansion}, truncated to keep only the $\nu^4$ auto- and cross-spectra. We note that $\mathbf{C}_\ell^{ET} = \left(\mathbf{C}_\ell^{TE}\right)^\mathsf{T}$, leaving the full covariance matrix positive definite.  This covariance matrix allows us to compute the Fisher information matrix defined in  Eq.~\eqref{eq:fisher_CMB}, where, in this case, the derivatives are taken with respect to the cross-spectra
\begin{align}
    C_\ell^{\mathrm{cross},TT} &\equiv C_{\ell}^{\Delta T_4 T}, \nonumber \\
    C_\ell^{\mathrm{cross},EE} &\equiv C_{\ell}^{\Delta E_4 E}, \nonumber \\
    C_\ell^{\mathrm{cross},TE} &\equiv C_{\ell}^{T \Delta E_4}, 
    \nonumber \\
    C_\ell^{\mathrm{cross},ET} &\equiv C_{\ell}^{E\Delta T_4}.
    \label{eq:def_cross}
\end{align}

The total signal-to-noise ratio of the cross correlation between the primary CMB and the Rayleigh scattering signal is then given by 
\begin{equation}
    \mathrm{SNR} = \left[ \sum_\ell S_\ell \cdot \mathbf{F}_\ell \cdot S_\ell^\mathsf{T} \right]^{1/2},
\end{equation}
where $S_\ell \equiv \begin{pmatrix}C_\ell^{\mathrm{cross},TT} & C_\ell^{\mathrm{cross},EE} & C_\ell^{\mathrm{cross},TE} & C_\ell^{\mathrm{cross},ET}\end{pmatrix}$ and $\mathbf{F}_\ell$ is defined in Eq.~\eqref{eq:fisher_CMB}.

\begin{figure*}[!ht]
\subfloat[]{
\includegraphics[width=0.475\linewidth]{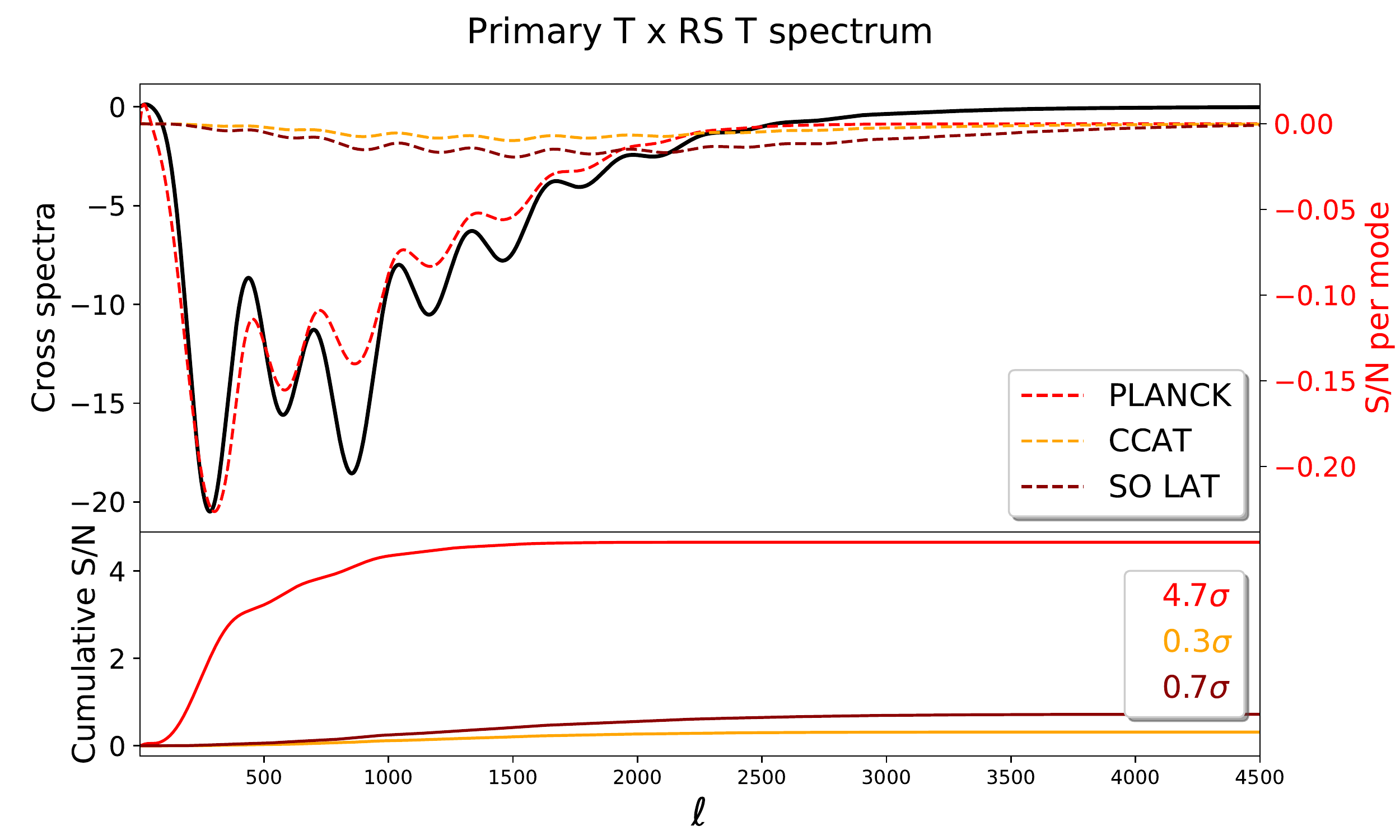}
}
\hfill
\subfloat[]{
\includegraphics[width=0.475\linewidth]{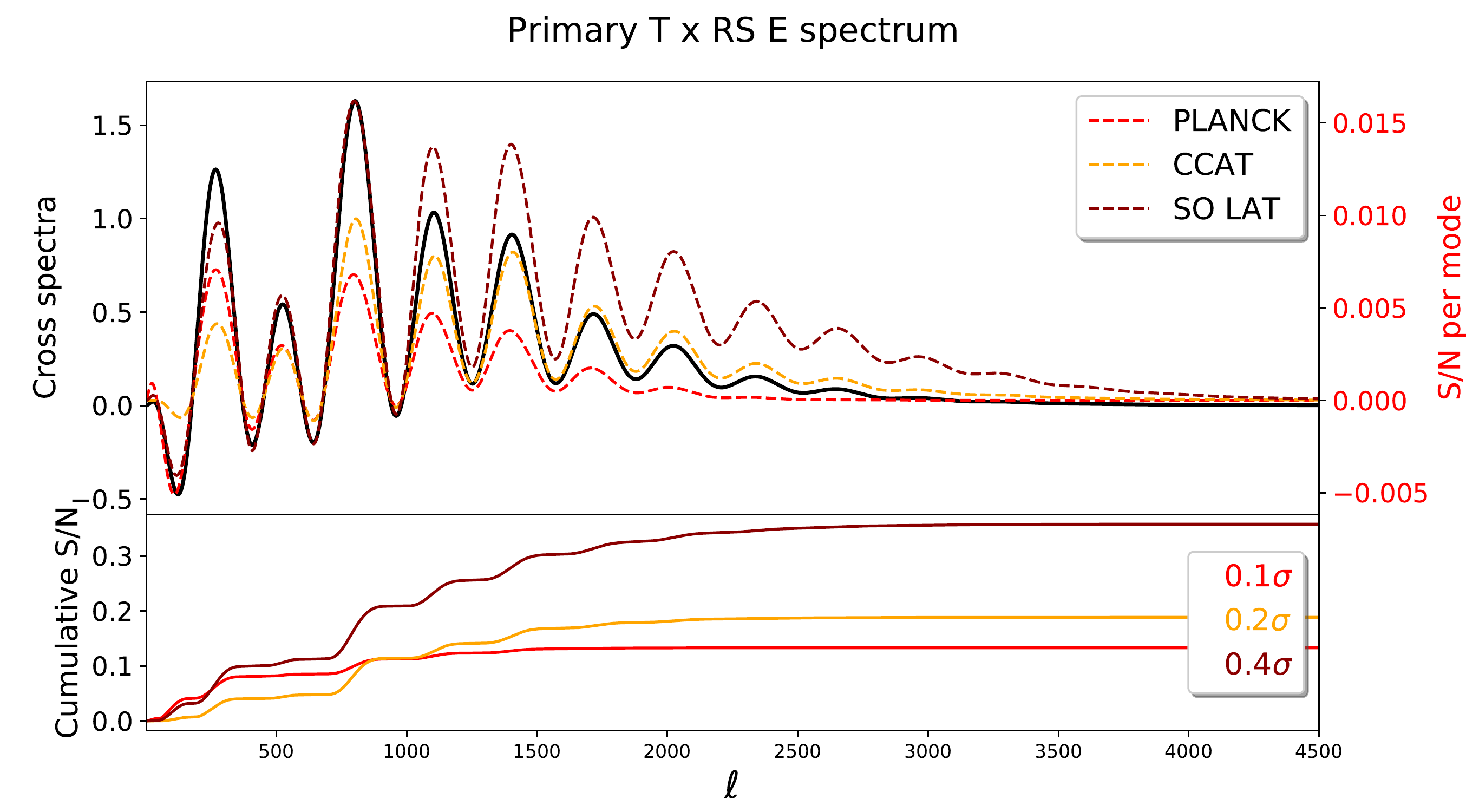}
}
\newline
\subfloat[]{
\includegraphics[width=0.475\linewidth]{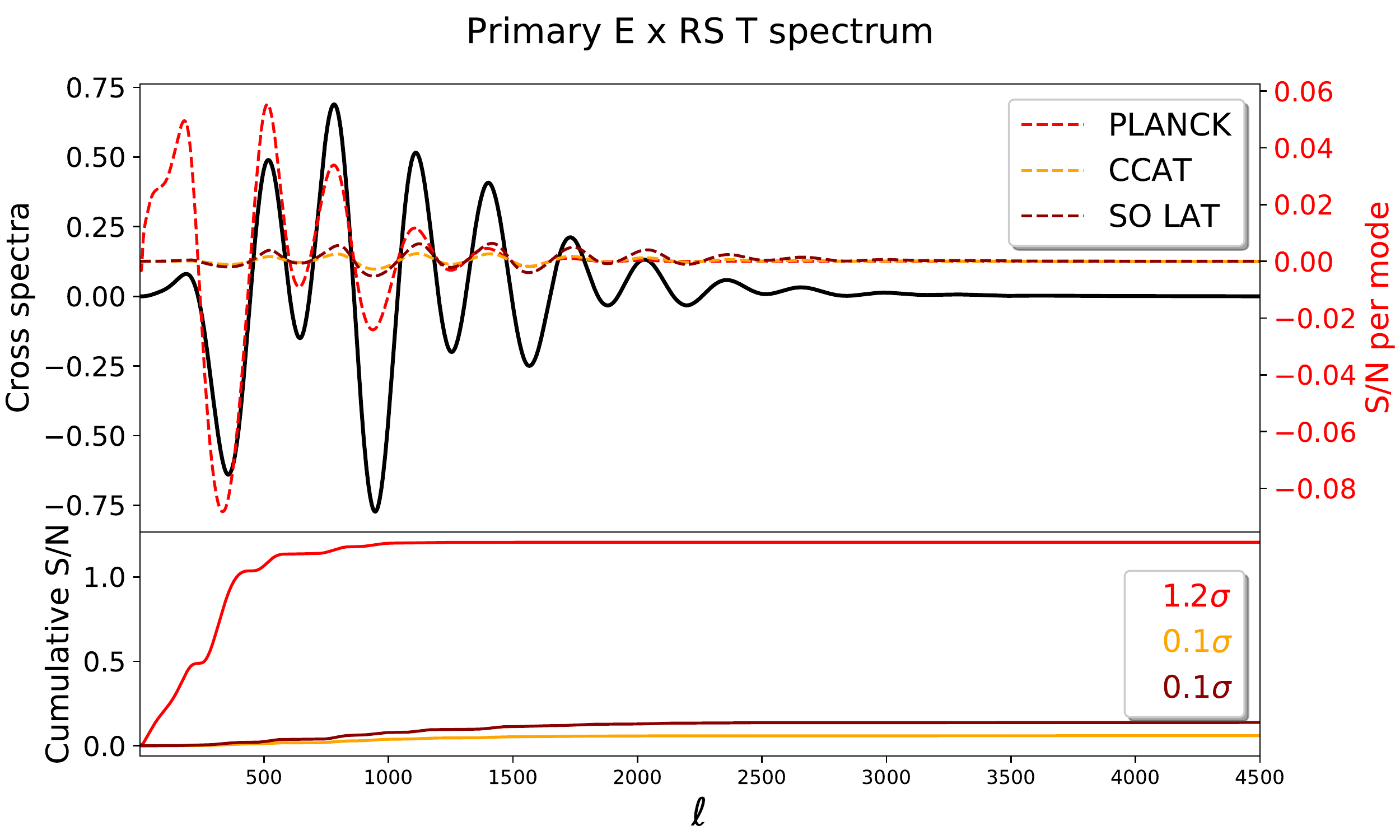}
}
\hfill
\subfloat[]{
\includegraphics[width=0.475\linewidth]{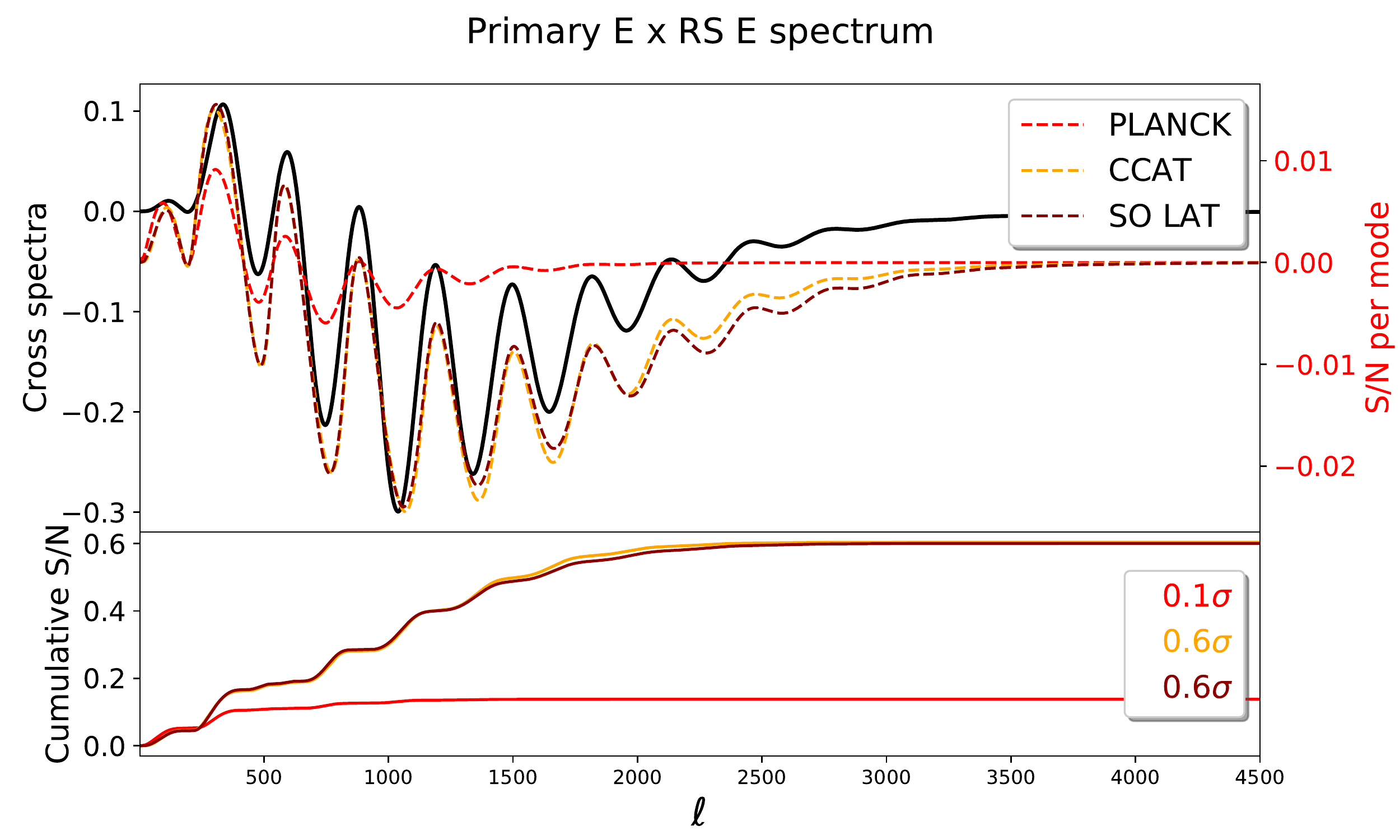}
}
\caption{Detectability of the 4 primary $\times$ Rayleigh cross power spectra for 3 experiments :  {\it Planck}  (red),  CCAT-prime (orange) and SO  large aperture telescope (dark red). Because of the large atmospheric noise on large scales, ground based experiment do not perform well in temperature. However, they provide a significant improvement over {\it Planck} in polarization.}
\label{fig:SN_CCAT_SO_PLANCK}
\end{figure*}

\subsubsection{Next generation CMB experiments}

Despite several experiments targeting CMB anisotropies both on larger scales from space ({\it Planck} satellite~\cite{Akrami:2018vks}) and on smaller scales from the ground (SPT~\cite{Austermann:2012ga}, ACT~\cite{Thornton:2016wjq}), there has not been a detection of cosmological Rayleigh scattering. {\it Planck} in principle has the sensitivity to detect the signal \cite{Lewis_2013}, but it is likely astrophysical foregrounds and the limited sensitivity in polarization has prevented a significant detection. The next generation of CMB experiments, which will be composed of both ground and space-based surveys, will have much better polarization sensitivity and their frequency coverage will allow a better treatment of astrophysical foregrounds on small scales. The experiments that we consider in this paper are CCAT-prime~\cite{CCAT19} and Simons Observatory (SO)~\cite{collaboration2019simons} which are two ground-based experiments. We have also included forecasts for a Stage-4 CMB experiment (CMB-S4)~\cite{Abazajian:2016yjj} as well as LiteBIRD~\cite{Hazumi2019} and PICO~\cite{PICO19} (see Appendix~\ref{app:experiments} for further details on these experiments). Forecasts for these experiments are summarised in Table~\ref{tab:SN_RS}. 

First of all, we note that space based telescopes are better suited to look for Rayleigh scattering, especially in temperature. LiteBIRD should improve on {\it Planck} by almost an order of magnitude, while PICO would have the necessary sensitivity and frequency coverage to significantly detect all four distinct primary-Rayleigh scattering cross-correlations. Ground-based experiments are heavily impacted by the atmosphere which hampers measurement of large scale fluctuations, especially in temperature, as observed in Fig.~\ref{fig:SN_CCAT_SO_PLANCK}. To assess what limits observations from the ground, we also present how a change in the number of detectors $N_{\rm det}$ and a lowering of the so-called $\ell_{\rm knee}$ affects the detectability for CCAT-prime. These quantities will affect the noise level directly (see Eq.~\eqref{eq:noise_model_atmo}). The white noise level $N_{\rm white}$ scales linearly with the number of detectors while $\ell_{\rm knee}$ controls the transition scale between the multipoles where the noise is dominated by the atmospheric contribution and the instrument contribution. In polarization, $N_{\rm red} = N_{\rm white}$ which means that a larger number of detectors in the focal plane (or equivalently a longer integration time) will improve measurements on all scales. $\ell_{\rm knee}$ is expected to decrease with altitude as the atmosphere becomes more dilute. Also, at the south pole, where SPT is located, a lower $\ell_{\rm knee}$ is observed~\cite{Choi20}). The CCAT-prime noise model is currently calibrated on available ACT measurements. However CCAT-prime will be located at a site 500~m above ACT, likely making our treatment of $\ell_\mathrm{knee}$ for CCAT-prime conservative.

While ground-based observations will most likely need to be combined with satellite observations to detect Rayleigh scattering, they will provide useful information in polarization. They will also map foregrounds on small scales which will help characterizing them and eventually mitigating their impact on cosmological analyses. Fig.~\ref{fig:SN_CCAT_SO_PLANCK} presents the Fisher forecasts for the four primary-Rayleigh cross power spectra for SO and CCAT-prime compared with {\it Planck} in the absence of foregrounds. The total signal-to-noise for {\it Planck}, CCAT-prime, and SO are $4.8\sigma$, $0.67\sigma$, and $0.97\sigma$ respectively. Combining the three experiments and properly accounting for overlapping sky coverage, yields a total signal-to-noise of $5.2\sigma$. This modest improvement from including ground based surveys could have a large impact as it comes with both some signal in polarization and improved measurements of foreground properties which is substantial in looking for a first detection of Rayleigh scattering. 
 
 \begin{table*}[]
    \centering
    \begin{tabular}{|cc|c|c|c|c|c|c|c|c|}
    \hline
         && Planck & SO LAT & CCAT-prime & CCAT-prime : $\ell_\mathrm{knee}/2$ & CCAT-prime : $2\times N_\mathrm{det}$ & CMB-S4 & LiteBIRD & PICO \\
    \hline \hline
     \multicolumn{2}{|c|}{$T_\mathrm{CMB} \times T_\mathrm{RS}$}   & 4.7  & 0.7  & 0.3  & 1.2 & 0.3 & 2.0 & 25 & 715 \\
    \hline
     \multicolumn{2}{|c|}{$E_\mathrm{CMB} \times E_\mathrm{RS}$}   & 0.1  & 0.6  & 0.6  & 0.7 & 0.9 & 1.8 & 1.4 & 45 \\
    \hline
     \multicolumn{2}{|c|}{$T_\mathrm{CMB} \times E_\mathrm{RS}$}   & 0.1 & 0.4  & 0.2  & 0.4 & 0.3 & 1.0 & 0.9 & 30 \\
    \hline
     \multicolumn{2}{|c|}{$E_\mathrm{CMB} \times T_\mathrm{RS}$}   & 1.2  & 0.1  & 0.1  & 0.2 & 0.1 & 0.4 & 10  & 195 \\
    \hline

    \end{tabular}
    \caption{Forecasted signal-to-noise ratio for detecting Rayleigh scattering for the four cross correlations for a set of CMB experiments. Ground-based experiments, although severely impacted by the effects of the earth's atmosphere when observing temperature anisotropies on large scales, will outperform {\it Planck} in polarization.}
    \label{tab:SN_RS}
\end{table*}

\subsection{Parameter forecasts}
In this section we forecast the expected improvements to parameter constraints that come from including the Rayleigh scattering signal for future CMB experiments. We start by presenting the methodology used as well as the assumed fiducial cosmology. 

\subsubsection{Forecast design}

Forecasts carried out it this paper closely follow the methodology described in~\cite{Wu14} in particular using the Fisher formalism defined in Sec.~\ref{subsubsec:fisher_formalism} where $\theta_i,\theta_j$ are the cosmological parameters of interest and $\mathbf{C}_\ell$ is the frequency-covariance matrix that includes auto- and cross-spectra both in temperature and polarization as well as the lensing potential spectrum $C_\ell^{\phi\phi}$. Since Rayleigh scattering is a frequency-dependent effect, all frequency auto- and cross-spectra have to be carefully accounted for. Spectra are modified according to Eq.~\eqref{eq:rs_expansion} keeping only $\nu^4$ terms. Therefore, the total covariance matrix reads:
\begin{equation}
    \mathbf{C}_\ell = \begin{pmatrix}
                      \mathbf{C}^{TT}_\ell + \mathbf{N}_\ell^{TT} & \mathbf{C}_\ell^{TE} & 0 \\
                      \mathbf{C}_\ell^{ET} & \mathbf{C}_\ell^{EE}+ \mathbf{N}_\ell^{EE} & 0 \\
                      0 & 0 & C_\ell^{\phi\phi} + N_\ell^{\phi\phi}\\
                      \end{pmatrix},
          \label{eq:cov_parameter}
\end{equation}
where $\mathbf{C}^{TT}_\ell$,  $\mathbf{C}^{TE}_\ell$,  $\mathbf{C}^{ET}_\ell$,  and $\mathbf{C}^{EE}_\ell$ are $N_\nu\times N_\nu$ frequency-covariance matrices. Similarly, noise power spectra $\mathbf{N}_\ell^{TT}$ and $\mathbf{N}_\ell^{EE}$ are $N_\nu\times N_\nu$ diagonal matrices defining the noise in temperature and polarization at each frequency. We will compare forecasts including Rayleigh scattering with ones including only the primary CMB. 
We forecast the primary-only constraints by creating a single effective channel in temperature and polarization with an inverse variance weighted noise given by
\be
N_\ell^{XX} &=& \left(\sum_{i =1}^{N}\left(N_{\ell,i}^{XX}\right)^{-1}\right)^{-1} \, . 
\ee
Where $N_{\ell,i}^{XX}$ are the noise spectra at each frequency. The error on the reconstructed gravitational potential, $N_\ell^{\phi \phi}$, is estimated following~\cite{Hu02} and its public implementation in \texttt{quicklens}\footnote{\url{ https://github.com/dhanson/quicklens}}. Following guidelines in~\cite{Wu14}, we will only use the $EB$ estimator since, for the noise levels considered, it provides the lowest noise estimates. \\

We consider constraints on the 6 parameters that define the $\Lambda$CDM cosmology.  We also show forecasts for three extensions to $\Lambda$CDM: changes to the light relic density parameterized by $N_\mathrm{eff}$, changes to the sum of neutrino masses $\sum m_\nu$,  and modifications to the primordial helium abundance $Y_\mathrm{He}$. Fiducial values are taken from~\cite{Akrami:2018vks} and summarized in Tab.~\ref{tab:fiducial_steps} together with the step size used for numerical derivatives.  Unless otherwise stated, forecasted errors are marginalized over the other parameters of the $\Lambda$CDM model. 

\begin{table*}[!ht]
    \centering
    \begin{tabular}{|c|c|c|}
    \hline
    Parameter & Fiducial value & Step size \\
    \hline \hline
    $\Omega_bh^2$ & $0.02237$ &  $8\times 10^{-4}$ \\
    \hline
    $\Omega_ch^2$ & $0.120$ & $3\times 10^{-3}$  \\
    \hline
    $10^9A_s$ & $2.099$ & $0.1$ \\
    \hline
    $n_s$ & $0.9649$ & $0.01$ \\
    \hline
    $H_0 \quad [\text{km/s/Mpc}]$ & $67.3$ & $0.5$ \\
    \hline
    $\tau$ & $0.0544$ & $0.02$ \\
    \hline
    $N_\mathrm{eff}$ & $3.046$ & $0.08$ \\
    \hline
    $\sum m_\nu \quad [\text{meV}]$  & $60$ & $10$   \\
    \hline
    $Y_\mathrm{He}$ & $0.2477$ & $0.04$ \\
    \hline
    \end{tabular}
    \caption{Fiducial values of the cosmological parameters considered in this analysis along with step sizes used for numerical derivatives. }
    \label{tab:fiducial_steps}
\end{table*}

When stated, we will combine our CMB experiment with baryon acoustic oscillation (BAO) constraints from a DESI-like galaxy survey~\cite{Font-Ribera:2013rwa}. This is done by simply adding the BAO Fisher matrix, $F^\mathrm{tot} = F^\mathrm{BAO} + F^\mathrm{CMB}$, since the observations are independent. This will improve constraints on the sum of the neutrino masses as well as the densities and the Hubble parameter. Details of BAO fisher matrix are given in Appendix~\ref{subsec:DESI}.

Furthermore, we use lensed CMB spectra (unless specified otherwise). Regarding the cut-off scale $\ell_{\rm max}$, we will first assume it is the same in polarization and temperature and take $\ell_{\rm max} = 5000$. This will allow a direct comparison with results from~\cite{Wu14}. Astrophysical foregrounds in temperature are significant at larger scales, and are comparable to the CMB power around $\ell \sim 3000$, so we show also forecasts which use $\ell_\mathrm{max}^{T}=3000$. In that case, we will also conservatively cut $TE$ spectra at $\ell_{\rm max} = 3000$.
We finally present forecasts for cosmic variance limited observations of the primary CMB on the full sky up to $\ell_\mathrm{max}=5000$ for comparison. 

\subsubsection{Results}

The only experiment we will consider for parameter forecasts is the proposed PICO satellite (see Appendix~\ref{subsec:PICO} for details of the experiment). Thanks to a broad and dense frequency coverage, PICO provides an example of an experiment which would greatly benefit from using Rayleigh scattering as an additional source of cosmological information. Results of our forecasts are presented in Tab.~\ref{tab:LCDM_forecasts} and \ref{tab:extensions_forecasts}. 

\begin{table*}[!ht]
    \hspace*{-1cm}\
    \centering
    \begin{tabular}{|cc||c|c|c|c|c|c||c|}
         \hline
         & &  $\Omega_bh^2$ & $\Omega_ch^2$ & $H_0 [\text{km/s/Mpc}]$ & $10^9A_s$ & $n_s$ & $\tau$ & Volume   \\ \hline \hline
          \multirow{4}{*}{Reference case} & PICO no Rayleigh & $2.30 \times 10^{-5}$ & $2.30 \times 10^{-4}$ & $8.78 \times 10^{-2}$ & $6.48 \times 10^{-3}$ & $1.27 \times 10^{-3}$ & $1.77 \times 10^{-3}$ & 1.0 \\\cline{2-9}
                                          & PICO with Rayleigh & $1.91 \times 10^{-5}$ & $2.14 \times 10^{-4}$ & $7.85 \times 10^{-2}$ & $6.07 \times 10^{-3}$ & $1.17 \times 10^{-3}$ & $1.67 \times 10^{-3}$ & 0.73 \\\cline{2-9}
                                          & Improvement & $16.98\%$ & $6.92\%$ & $10.52\%$ & $6.36\%$ & $7.61\%$ & $5.51\%$ & $28\%$ \\\cline{2-9}
                                          & Primary-only CVL & $7.94 \times 10^{-6}$ & $1.61 \times 10^{-4}$ & $6.06 \times 10^{-2}$ & $5.21 \times 10^{-3}$ & $7.02 \times 10^{-4}$ & $1.43\times10^{-3}$ & $9.5\times10^{-3}$\\\hline \hline
          \multirow{4}{*}{$\ell_\mathrm{max}^{T} = 3000$} & PICO no Rayleigh & $2.43 \times 10^{-5}$ & $2.38 \times 10^{-4}$ & $9.09 \times 10^{-2}$ & $6.50 \times 10^{-3}$ & $1.32 \times 10^{-3}$ & $1.77 \times 10^{-3}$ & 1.82 \\\cline{2-9}
                                          & PICO with Rayleigh & $2.00 \times 10^{-5}$ & $2.22 \times 10^{-4}$ & $8.10 \times 10^{-2}$ & $6.08 \times 10^{-3}$ & $1.21 \times 10^{-3}$ & $1.68 \times 10^{-3}$ & 1.29 \\\cline{2-9}
                                          & Improvement & $17.70\%$ & $6.99\%$ & $10.92\%$ & $6.43\%$ & $8.23\%$ & $5.45\%$ & $29\%$ \\\cline{2-9}
                                          & Primary-only CVL & $1.10 \times 10^{-5}$ & $1.69 \times 10^{-4}$ & $6.23 \times 10^{-2}$ & $5.28 \times 10^{-3}$ & $9.08 \times 10^{-4}$ & $1.46 \times 10^{-3}$ & $3.1\times10^{-2}$ \\\hline \hline
          \multirow{4}{*}{With DESI BAO} & PICO no Rayleigh & $2.30 \times 10^{-5}$ & $1.91 \times 10^{-4}$ & $7.29 \times 10^{-2}$ & $5.87 \times 10^{-3}$ & $1.21 \times 10^{-3}$ & $1.56 \times 10^{-3}$ & 0.8 \\\cline{2-9}
                                          & PICO with Rayleigh & $1.90 \times 10^{-5}$ & $1.82 \times 10^{-4}$ & $6.66 \times 10^{-2}$ & $5.60 \times 10^{-3}$ & $1.10 \times 10^{-3}$ & $1.50 \times 10^{-3}$ & 0.61 \\\cline{2-9}
                                          & Improvement & $17.51\%$ & $4.73\%$ & $8.62\%$ & $4.64\%$ & $8.80\%$ & $3.52\%$ & $26\%$ \\\cline{2-9}
                                          & Primary-only CVL & $7.89 \times 10^{-6}$ & $1.45 \times 10^{-4}$ & $5.45 \times 10^{-2}$ & $4.81 \times 10^{-3}$ & $6.67 \times 10^{-4}$ & $1.31 \times 10^{-3}$ & $8.5\times10^{-3}$ \\\hline \hline
          \multirow{4}{*}{With unlensed spectra} & PICO no Rayleigh & $1.98 \times 10^{-5}$ & $2.31 \times 10^{-4}$ & $8.59 \times 10^{-2}$ & $6.52 \times 10^{-3}$ & $1.23 \times 10^{-3}$ & $1.73 \times 10^{-3}$ & 0.67 \\\cline{2-9}
                                          & PICO with Rayleigh & $1.71 \times 10^{-5}$ & $2.15 \times 10^{-4}$ & $7.77 \times 10^{-2}$ & $6.12 \times 10^{-3}$ & $1.12 \times 10^{-3}$ & $1.65 \times 10^{-3}$ & 0.51 \\\cline{2-9}
                                          & Improvement & $13.76\%$ & $6.85\%$ & $9.57\%$ & $6.15\%$ & $8.31\%$ & $4.96\%$ & $24\%$\\\cline{2-9}
                                          & Primary-only CVL & $6.62 \times 10^{-6}$ & $1.48 \times 10^{-4}$ & $5.38 \times 10^{-2}$ & $5.02 \times 10^{-3}$ & $6.51 \times 10^{-4}$ & $1.35 \times 10^{-3}$ & $3.1\times10^{-3}$ \\\hline \hline
          \multirow{4}{*}{Fixing $Y_\mathrm{He}$ by BBN} & PICO no Rayleigh & $2.31 \times 10^{-5}$ & $2.30 \times 10^{-4}$ & $8.79 \times 10^{-2}$ & $6.49 \times 10^{-3}$ & $1.26 \times 10^{-3}$ & $1.77 \times 10^{-3}$ & 1.0 \\\cline{2-9}
                                          & PICO with Rayleigh & $1.92 \times 10^{-5}$ & $2.14 \times 10^{-4}$ & $7.85 \times 10^{-2}$ & $6.07 \times 10^{-3}$ & $1.17 \times 10^{-3}$ & $1.67 \times 10^{-3}$ & 0.73 \\\cline{2-9}
                                          & Improvement & $16.84\%$ & $7.07\%$ & $10.71\%$ & $6.45\%$ & $7.54\%$ & $5.62\%$ & $27\%$ \\\cline{2-9}
                                          & Primary-only CVL & $8.00 \times 10^{-6}$ & $1.61 \times 10^{-4}$ & $6.07 \times 10^{-2}$ & $5.22 \times 10^{-3}$ & $7.01 \times 10^{-4}$ & $1.44 \times 10^{-3}$ & 0.01 \\\hline \hline
          \multirow{4}{*}{Not including $C_\ell^{\phi \phi}$} & PICO no Rayleigh & $2.39 \times 10^{-5}$ & $2.83 \times 10^{-4}$ & $1.08 \times 10^{-1}$ & $6.62 \times 10^{-3}$ & $1.40 \times 10^{-3}$ & $1.77 \times 10^{-3}$ & 1.79 \\\cline{2-9}
                                          & PICO with Rayleigh & $1.93 \times 10^{-5}$ & $2.48 \times 10^{-4}$ & $9.09 \times 10^{-2}$ & $6.34 \times 10^{-3}$ & $1.31 \times 10^{-3}$ & $1.68 \times 10^{-3}$ & 1.21 \\\cline{2-9}
                                          & Improvement & $19.29\%$ & $12.28\%$ & $16.11\%$ & $4.24\%$ & $6.69\%$ & $5.28\%$ & $33\%$\\\cline{2-9}
                                          & Primary-only CVL & $1.02 \times 10^{-5}$ & $1.75 \times 10^{-4}$ & $6.57 \times 10^{-2}$ & $5.35 \times 10^{-3}$ & $9.88 \times 10^{-4}$ & $1.45 \times 10^{-3}$ & 0.06  \\\hline

    \end{tabular}
    \caption{1-$\sigma$ errors on $\Lambda$CDM parameters. Errors are quoted as marginalized over the remaining parameters. Parameter volume is normalized to the reference case and computed as the square root of the determinant of the parameter-covariance matrix. }
    \label{tab:LCDM_forecasts}
\end{table*}

\begin{table*}[!ht]
    \centering
    \begin{tabular}{|cc||c|c|c|}
         \hline
         & &  $N_\mathrm{eff}$ & $\sum m_\nu [\text{meV}]$ & $Y_\mathrm{He}$   \\\hline \hline
          \multirow{4}{*}{Reference case} & PICO no Rayleigh & $3.06 \times 10^{-2}$ & $35.9$ & $1.67 \times 10^{-3}$ \\\cline{2-5}
                                          & PICO with Rayleigh & $2.81 \times 10^{-2}$ & $16.7 $ & $1.44 \times 10^{-3}$  \\\cline{2-5}
                                          & Improvement & $8.21\%$ & $53.55\%$ & $13.75\%$ \\\cline{2-5}
                                          & Primary-only CVL & $9.68 \times 10^{-3}$ & $21.0 $ & $5.83 \times 10^{-4}$ \\\hline \hline
          \multirow{4}{*}{$\ell_{\rm max}^{T} = 3000$} & PICO no Rayleigh & $3.60 \times 10^{-2}$ & $36.6$ & $2.08 \times 10^{-3}$  \\\cline{2-5}
                                          & PICO with Rayleigh & $3.24 \times 10^{-2}$ & $16.7 $ & $1.73 \times 10^{-3}$ \\\cline{2-5}
                                          & Improvement & $9.96\%$ & $54.27\%$ & $16.76\%$  \\\cline{2-5}
                                          & Primary-only CVL & $1.82 \times 10^{-2}$ & $27.7 $ & $1.13 \times 10^{-3}$  \\\hline \hline
          \multirow{4}{*}{With DESI BAO} & PICO no Rayleigh & $2.85 \times 10^{-2}$ & $11.5$ & $1.65 \times 10^{-3}$ \\\cline{2-5}
                                          & PICO with Rayleigh & $2.63 \times 10^{-2}$ & $10.1$ & $1.41 \times 10^{-3}$ \\\cline{2-5}
                                          & Improvement & $7.72\%$ & $11.85\%$ & $14.56\%$ \\\cline{2-5}
                                          & Primary-only CVL & $9.47 \times 10^{-3}$ & $9.70$ & $5.79 \times 10^{-4}$ \\\hline \hline
          \multirow{4}{*}{With unlensed spectra} & PICO no Rayleigh & $2.41 \times 10^{-2}$ & $33.8$ & $1.32 \times 10^{-3}$   \\\cline{2-5}
                                          & PICO with Rayleigh & $2.28 \times 10^{-2}$ & $16.4$ & $1.25 \times 10^{-3}$ \\\cline{2-5}
                                          & Improvement & $5.15\%$ & $51.48\%$ & $5.65\%$  \\\cline{2-5}
                                          & Primary-only CVL & $8.04 \times 10^{-3}$ & $17.3$ & $4.84 \times 10^{-4}$ \\\hline \hline
          \multirow{4}{*}{Fixing $Y_\mathrm{He}$ by BBN} & PICO no Rayleigh & $2.48 \times 10^{-2}$ & $35.9$  & .. \\\cline{2-5}
                                          & PICO with Rayleigh & $2.31 \times 10^{-2}$ & $16.1$ & .. \\\cline{2-5}
                                          & Improvement & $6.85\%$ & $55.00\%$ & ..\\\cline{2-5}
                                          & Primary-only CVL & $7.88 \times 10^{-3}$ & $21.0$ & .. \\\hline \hline
          \multirow{4}{*}{Not including $C_\ell^{\phi \phi}$} & PICO no Rayleigh & $3.50 \times 10^{-2}$ & $37.8$ & $1.91 \times 10^{-3}$  \\\cline{2-5}
                                          & PICO with Rayleigh & $3.04 \times 10^{-2}$ & $17.4$ & $1.49 \times 10^{-3}$  \\\cline{2-5}
                                          & Improvement & $13.05\%$ & $53.84\%$ & $22.03\%$ \\\cline{2-5}
                                          & Primary-only CVL & $1.46 \times 10^{-2}$ & $26.3$ & $8.50 \times 10^{-4}$  \\\hline

    \end{tabular}
    \hspace*{-1cm}
    \caption{1-$\sigma$ errors on 3 extensions of $\Lambda$CDM. Errors are quoted are marginalized over $\Lambda$CDM, keeping the two other parameters fixed at their fiducial value.}
    \label{tab:extensions_forecasts}
\end{table*}

\begin{figure*}[t]
    \centering
    \includegraphics[width = \textwidth]{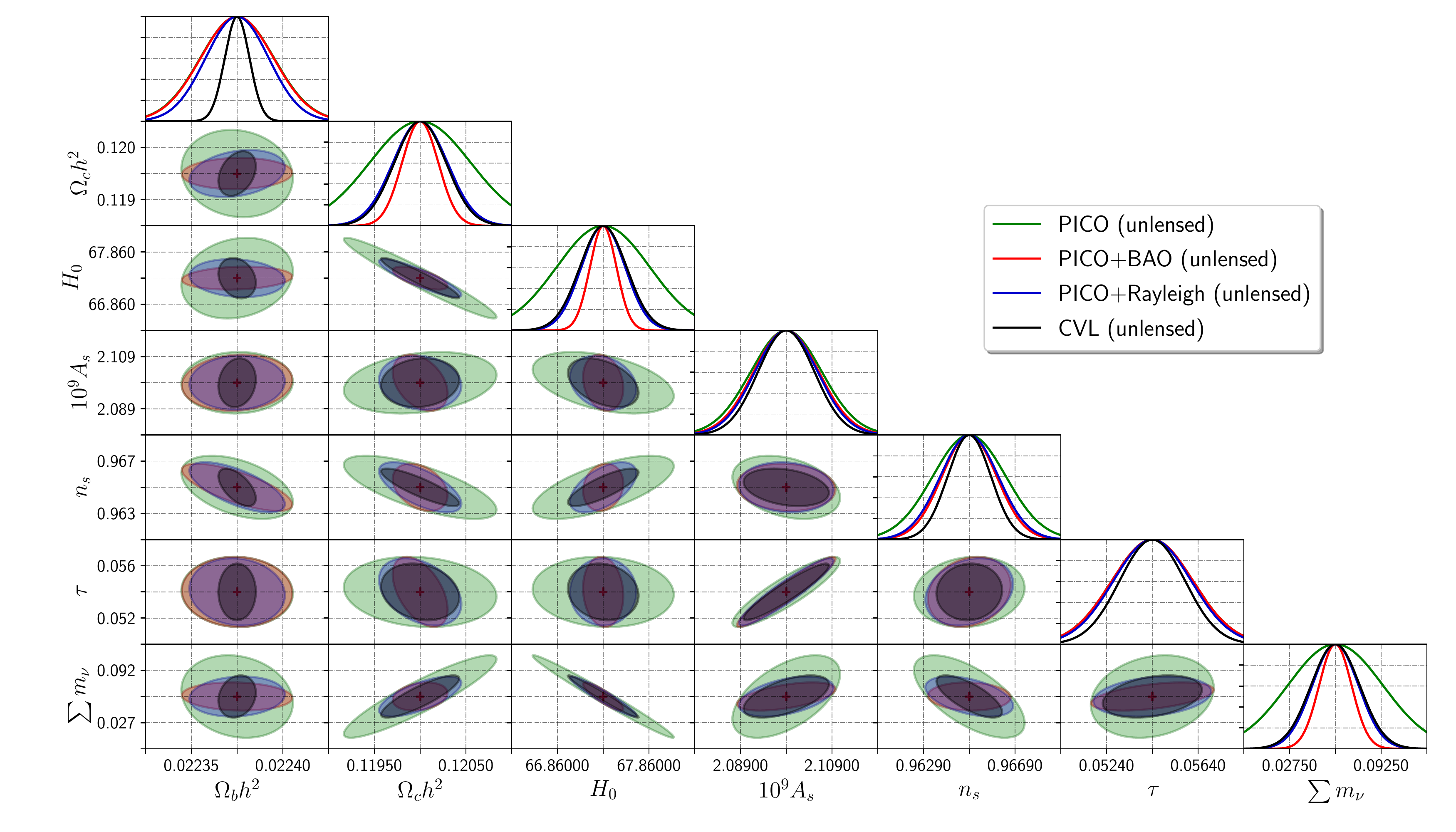}
    \caption{1-$\sigma$ contours for the Fisher forecasts for the $\Lambda$CDM+$\sum m_\nu$ cosmology. We are using unlensed spectra,  including $C_\ell^{\phi\phi}$, with a cut-off scale of $5000$. This corresponds to the fourth line of Tab.~\ref{tab:extensions_forecasts}. 
    }
    \label{fig:triangular_plot}
\end{figure*}

\paragraph{$\Lambda$CDM forecasts} 
As discussed in Sec.~\ref{sec:cosmo_parameters}, Rayleigh scattering can be particularly useful to constrain parameters that affect a physical distance scale in the early Universe. Indeed, this distance scale will project to different angular scales at last scattering for different frequencies. A first example is the comoving size of the sound horizon which is affected by baryon and dark matter densities $\Omega_b h^2$ and $\Omega_c h^2$. A better measurement of the sound horizon would also provide a better measurement of the Hubble parameter $H_0$. Our forecasts show improved constraints on these parameters in the first 3 columns of Tab.~\ref{tab:LCDM_forecasts}. The improvement on the last three parameters is attributed to the breaking of degeneracies made possible by including Rayleigh scattering (for example, we can observe the $n_s$ contours changing when including Rayleigh information in Fig.~\ref{fig:triangular_plot}). We also observe a reduction of the parameter volume, defined as $ \propto \left[{\rm det} \left({\bf F}^{-1}\right)\right]^{1/2}$ where $ {\bf F}$ is the Fisher matrix defined in Eq.~\ref{eq:fisher_tot} . Typically, including Rayleigh scattering to the reference case provides a similar reduction to the addition of BAO information from DESI. \\

\begin{figure}[!h]
    \centering
    \includegraphics[width =\columnwidth]{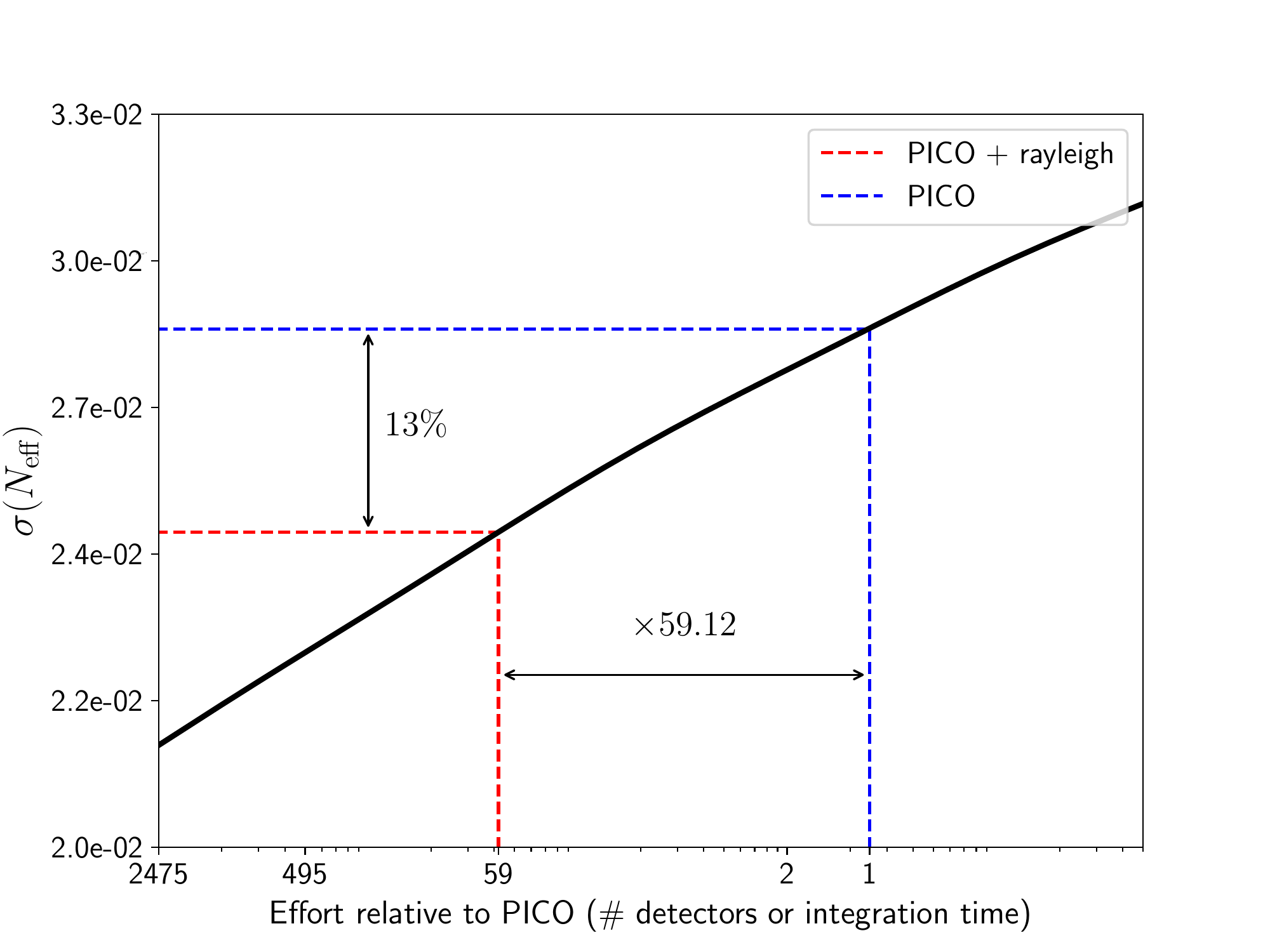}
    \caption{Forecasted 1-$\sigma$ error on $N_{\rm eff}$ as a function of the effort (as measured in detector-years) by a space-based CMB experiment with a $5'$ beam, normalized such that an effort of 1 provides similar results as PICO, not including Rayleigh scattering. These forecasts use lensed CMB spectra and do not include constraints derived from lensing reconstruction. $Y_{\rm He}$ is fixed by BBN consistency. To obtain a similar $\sim 10\%$ improvement without including constraints from Rayleigh scattering, we would need 60 times more effort.}
    \label{fig:Neff_effort}
\end{figure}

\begin{figure}[t]
    \centering
    \includegraphics[width= .48\textwidth]{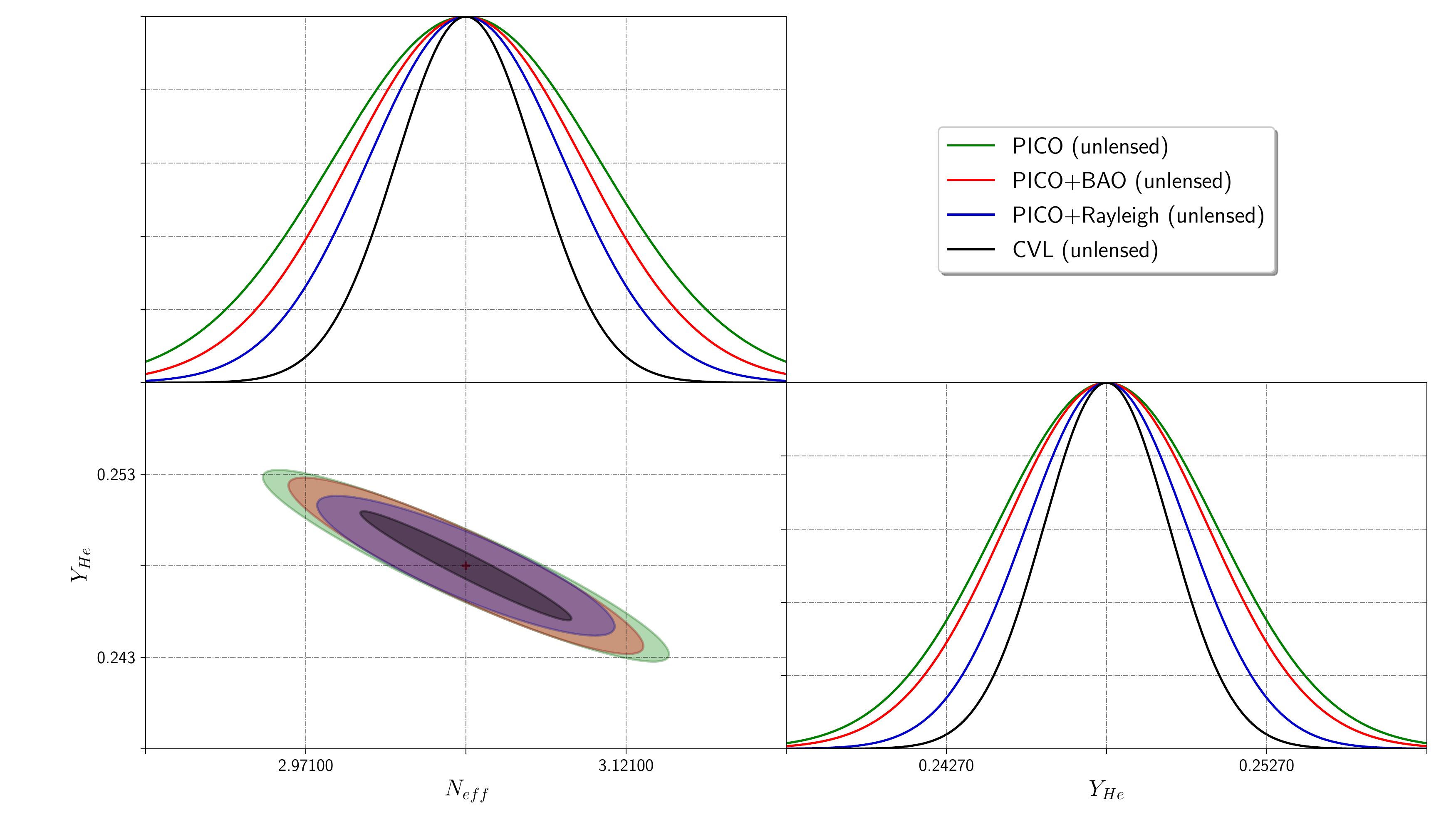}
    \caption{1-$\sigma$ contours for the Fisher forecasts in the $Y_\mathrm{He}, N_\mathrm{eff}$ plane. We are using unlensed spectra, including $C_\ell^{\phi\phi}$, with a cut-off scale of $5000$. This corresponds to the fourth line of Tab.~\ref{tab:extensions_forecasts}. }
    \label{fig:YHe_Neff}
\end{figure}

\paragraph{Forecasts including extensions to $\Lambda$CDM} We have considered three extensions to $\Lambda$CDM : $N_\mathrm{eff}$, $Y_\mathrm{He}$, and $\sum m_\nu$. As expected from Sec.~\ref{sec:cosmo_parameters}, the primordial helium abundance is directly probed by the amplitude of the Rayleigh scattering signal. $N_\mathrm{eff}$ constraints are improved in two ways. First, when both $Y_\mathrm{He}$ and $N_\mathrm{eff}$ are allowed to vary, the measurement of Rayleigh scattering mitigates the degenereacy in the $Y_\mathrm{He}$-$N_\mathrm{eff}$ plane  (Fig.~\ref{fig:YHe_Neff}).  However, the two parameters retain a significant correlation due to their similar effect on the damping scale, and any improvement on the measurement of $Y_\mathrm{He}$ would yield a better measurement of $N_\mathrm{eff}$. However, the improvement to the constraint on $N_\mathrm{eff}$ persists when the Helium fraction is fixed to be consistent with the predictions of Big Bang Nucleosynthesis (BBN). Similar to the size of the sound horizon, the damping scale provides a fixed physical length scale which will project onto different angular scales at different frequencies thanks to Rayleigh scattering, explaining the improved constraints on $N_\mathrm{eff}$ even when $Y_\mathrm{He}$ is fixed by BBN consistency. Although the improvement of $\sim 10\%$ might seem modest, measuring $N_{\rm eff}$ with the primary CMB will become more challenging with future experiments as constraints rely on the small scale CMB. Due to the exponential damping of the power spectrum on small scales, new small scale modes are difficult to measure and to disentangle from foregrounds especially from space where the size of the telescope (which controls the beam size) is limited. In Fig~.\ref{fig:Neff_effort}, we show that to reach $\sim 10\%$ improvement on $\sigma(N_{\rm eff})$ without Rayleigh scattering would require roughly $50$ times more detectors for a PICO-like mission which is unlikely to happen in practice. 

Finally, CMB measurement of the sum of neutrino masses $\sum m_\nu$ benefits greatly from including Rayleigh scattering. The improvement on $\sum m_\nu$ comes from the fact that Rayleigh scattering allows for a tighter constraint on $\Omega_c h^2$ (as seen in Fig.~\ref{fig:triangular_plot}). The effects of $\Omega_c h^2$ and $\sum m_\nu$ on the CMB lensing power spectrum are similar which causes these parameters to exhibit a degeneracy (that is typically assumed to be broken by including external BAO data), but the degeneracy is also broken when Rayleigh scattering information is included. This improvement from Rayleigh scattering allows for a high significance measurement of even the minimal sum of neutrino masses which solely relies on CMB data. 

\section{Discussion and Conclusions}

Rayleigh scattering of the cosmic microwave background is an interesting source of additional cosmological information. By increasing the comoving opacity in a frequency-dependent way around recombination it generates a unique signature and is responsible for distinctive distortions to the primary power spectra. On large angular scales, $E$-mode polarization anisotropies are boosted whereas on small scales, both temperature and polarization signals are damped in a frequency-dependent way. The redshift of last scattering also becomes frequency dependent. We have shown that these two effects provide complementary information to the primary CMB which can improve cosmological parameter constraints. The heuristic argument for this improvement is that any distance scale in the early Universe will be projected onto different angular scales at each frequency by Rayleigh scattering due to the frequency dependence of the visibility function. Interestingly, this can lead to promising improvements on some key parameters. We found that $N_\mathrm{eff}$ is improved by $\sim 10\%$ for a PICO-like experiment. Most of the constraining power on $N_{\rm eff}$ in primary spectra comes from small scales that are exponentially damped and obscured by astrophysical foregrounds. Furthermore, space mission quickly run out of accessible small scales modes because of their beam size. By allowing constraints on $N_{\rm eff}$ to not solely rely on these small scales, Rayleigh scattering opens a new window for improvements that would be challenging to match using primary spectra only.

Similarly, the constraint on $\sum m_\nu$ is also improved by including Rayleigh scattering, and a PICO-like experiment should be able to measure the minimal sum of neutrino masses at almost $4\sigma$ without relying on external datasets. This would make the understanding and treatment of systematics easier since only one experiment will be involved. It is also worth noting that these improvements come for free, as no modifications to these experiments are required to measure the signal as long as enough high frequency channels are available and foregrounds remain under control.

Astrophysical foregrounds have been neglected throughout this paper. While they are likely to be one of the reasons why {\it Planck} has not been able to detect Rayleigh scattering, the next generation of ground-based experiments will provide more data that will help understand foregrounds on small scales and at higher frequencies. These experiments will also produce improved measurements of polarization anisotropies which are less prone to foreground contamination. In future work, we plan to carefully study the impact of foregrounds on detectability forecasts. Especially, it will be important to decide which component separation method is the most suitable to detect the Rayleigh scattering signal. Although non-parametric methods such as ILC (or variation of it) are appreciated for their blindness they might not be well suited for this particular case. Indeed, the fact that the largest signal is the primary $\times$ Rayleigh cross-spectrum means that, in practice, two maps will be required: one that contains only the primary CMB and the other one that contains only the Rayleigh scattering signal. In order to obtain an unbiased estimate of the cross-correlation, the Rayleigh signal and the primary CMB will need to be deprojected which comes at the price of an increased variance in the cleaned maps~\cite{Remazeilles11}. Rayleigh scattering benefits from the fact it can be completely and accurately modelled once the cosmological parameters are fixed. Hence, Rayleigh scattering might be better suited to parametric cleaning methods where each component in the sky would be modelled and fitted for. 

In this paper we show the Rayleigh signal has the potential to become a true cosmological observable in the future. Since the signal itself is guaranteed, this should provide strong motivation to include it as a target in future cosmological surveys. Although the amplitude of the induced distortion is small, and the study presented here does not include foregrounds or systematic effects,
lower noise and improved frequency coverage should lead to a detection in the relatively near future, opening up a new window into the early Universe.

\section*{Acknowledgments}

The authors would like to thank Bradford Benson, Jean-François Cardoso, Steve Choi, Will Coulton, Antony Lewis, Mike Niemack and Yijie Zhu for useful discussions. BB is supported by the Science and Technologies Facilities Council. JM is supported by the US Department of Energy under grant no.~DE-SC0010129. P.D.M.\ acknowledges support of the Netherlands organization for scientific research (NWO) VIDI grant (dossier 639.042.730).

\appendix

\section{Deprojection of the correlated Rayleigh scattering signal.}
\label{app:eigenvalues}

From Fig.~\ref{fig:correlation}, we observe the Rayleigh scattering signal to be highly correlated with the primary CMB. In this Appendix we will investigate methods to deproject the Rayleigh scattering part of the signal from the primary one, following a  framework similar to the one presented in \cite{Scott_2016} in the context of temperature and $E$-mode polarization primary CMB signals. We will comment on how this affects the detectability of Rayleigh scattering.

Naively, Rayleigh scattering is modelled as a frequency- and scale-dependent distortion (Eq.~\ref{eq:signal_rs}). This leads to non-zero cross-correlation between the different terms in Eq.~\eqref{eq:signal_rs}, which can be rewritten for the temperature case as: 
\begin{equation}
    \mathbf{a}_{\ell m} = \mathbf{A}_\ell \begin{pmatrix} 
    a_{\ell m}^T \\ 
    \Delta a_{\ell m} ^{T,4} 
    \end{pmatrix}    
    ,\text{with } \mathbf{A}_\ell \equiv 
    \begin{pmatrix}
    1 & \left( \frac{\nu_1}{\nu_0} \right)^4 \\
     & \vdots \\
    1 & \left( \frac{\nu_N}{\nu_0} \right)^4 \\
    \end{pmatrix},\label{eq:signal_freq}
\end{equation}
where $\mathbf{a}_{\ell m}$ is a $N_\nu$ dimensional vector and $N_\nu$ the number of frequencies in the experiment. The frequency-covariance matrix can also be written as a function of $\mathbf{A}_\ell$ (we will limit ourselves to temperature spectra): 
\begin{align}
    \mathbf{C}_\ell^{\nu\nu}     &\equiv \langle \mathbf{a}_{\ell m} \mathbf{a}_{\ell m'}^\dagger \rangle = \mathbf{A}_\ell \begin{pmatrix}
    C_\ell^{TT} & C_\ell ^{T\Delta T_4} \\
    C_\ell ^ {\Delta T_4T} & C_\ell^{\Delta T_4 \Delta T_4}
    \end{pmatrix}
    \mathbf{A}_\ell^t \nonumber \\
    &= \mathbf{A}_\ell \begin{pmatrix}
    C_\ell^0 & C_\ell ^{\rm cross} \\
    C_\ell ^ {\rm cross} & C_\ell^{\rm auto} 
    \end{pmatrix}
    \mathbf{A}_\ell^t
    = \mathbf{A}_\ell \mathbf{C}_\ell^s
    \mathbf{A}_\ell^t.
    \label{eq:freq_covmat}
\end{align}
Here we have defined ${\bf C}_\ell^s$ as the signal-covariance matrix.
Detecting the Rayleigh scattering signal can be made in two different ways: i) detecting its cross correlation with the primary CMB or ii) detecting its auto-spectra. In the presence of foregrounds we first need to clean the data. This can be achieved using an Internal Linear Combination (ILC), (see e.g. \cite{Remazeilles11}). ILCs are based on obtaining  a set of weights to apply at each frequency map $a_{\ell m}$ such that the resulting linear combination has minimum variance while preserving the signal of interest. The resulting noise for each signal is given by 
\begin{equation}
    \mathbf{n}_\ell \equiv \left[\mathbf{A}_\ell^t \mathbf{N}_\ell^{-1} \mathbf{A}_\ell\right]^{-1}
    \label{eq:ILC_noise}
\end{equation}
where $\mathbf{N}_\ell$ is the noise covariance matrix. In this work, we have neglected foregrounds, meaning that $\mathbf{N}_\ell$ only captures instrumental noise (and atmospheric noise for ground-based experiments). 

The detectability of Rayleigh scattering can be assessed by introducing a parameter $\alpha$ in front of the Rayleigh term $\Delta a_{\ell m}^{T,4}$ in Eq.~\eqref{eq:signal_freq}. We equate our ability to detect the Rayleigh signal to distinguishing $\alpha$ from 1. Following the Fisher formalism introduced in \ref{subsubsec:fisher_formalism}, the signal-to-noise of detecting Rayleigh scattering is given by:
\begin{equation}
    \mathbf{F}_\ell = f_{\rm sky}\frac{2\ell + 1}{2} {\rm Tr}\left[(\mathbf{C}_\ell^{s} + \mathbf{n}_\ell)^{-1} \frac{\partial\mathbf{C}_\ell^{s}}{\partial \alpha} (\mathbf{C}_\ell^{s} + \mathbf{n}_\ell)^{-1} \frac{\partial\mathbf{C}_\ell^{s}}{\partial \alpha} 
    \right],
\end{equation}
\begin{equation}
    {\rm SNR} = \left[\sum_\ell{\bf F}_\ell \right]^\frac{1}{2}
    \label{eq:SNR_alpha}
\end{equation}

The signal-to-noise of Rayleigh scattering is sourced by both the auto- and cross-spectra, with the latter being the largest contribution for the experiments considered. However, detecting cross-correlations might be challenging, since the ILC procedure will leave some residual CMB in the Rayleigh scattering map and vice-versa. The cross- correlation will then be biased by these residuals. To overcome this, one can add an additional constraint to the ILC by requiring a specific component to be deprojected \cite{Remazeilles11}. This comes at the cost of an increase in variance but removes the bias from residual contamination. 

Instead of using the physical but correlated signals $a_{\ell m}^T$ and $\Delta a_{\ell m}^{T,4}$, one can look for a different basis spanned by orthogonal vectors (ie. uncorrelated signals). This will result in a signal-covariance matrix ${\bf C}_\ell^{r}$ that is diagonal. This procedure has been explored in the context of primary CMB temperature and polarization signals and has been described in detail in~\cite{Scott_2016}. We will follow this  framework to de-correlate the primary CMB from the Rayleigh scattering signal. 

We first define a rotation of the signals basis by 
\begin{equation}
    \begin{pmatrix}
    a_{\ell m}^a \\
    a_{\ell m}^b 
    \end{pmatrix} = 
    \begin{pmatrix}
    \cos \theta_\ell & \sin \theta_\ell \\
    -\sin \theta_\ell & \cos \theta_\ell 
    \end{pmatrix} \cdot 
    \begin{pmatrix}
    a_{\ell m}^T\\
    \Delta a_{\ell m}^{T,4}
    \end{pmatrix}
    \label{eq:rotation}
\end{equation}
Imposing the constraint that $\langle a_{\ell m}^a a_{\ell m'}^b\rangle = 0$, yields: $\tan 2\theta_\ell = \frac{2C_\ell^{\bf cross}}{C_\ell^0 - C_\ell^{\bf auto}}$. The auto and cross-spectra then read
\begin{widetext}

\begin{align}
    \lambda_\ell^{aa} &\equiv \langle a_{\ell m}^a a_{\ell m'}^a \rangle  = \cos^2\theta_\ell C_\ell^0 + \sin^2\theta_\ell C_\ell^{\rm auto} + \sin 2\theta_\ell C_\ell^{\rm cross} = \frac{1}{2}\left(C_\ell^0 + C_\ell^{\rm auto}\right) + \frac{1}{2}\sqrt{\left(C_\ell^0 + C_\ell^{\rm auto}\right)^2 + 4\left(C_\ell^{\rm cross}\right)^2}, \\
    \lambda_\ell^{bb} &\equiv \langle a_{\ell m}^b a_{\ell m'}^b \rangle  = \sin^2\theta_\ell C_\ell^0 + \cos^2\theta_\ell C_\ell^{\rm auto} - \sin 2\theta_\ell C_\ell^{\rm cross} = \frac{1}{2}\left(C_\ell^0 + C_\ell^{\rm auto}\right) -  \frac{1}{2}\sqrt{\left(C_\ell^0 + C_\ell^{\rm auto}\right)^2 + 4\left(C_\ell^{\rm cross}\right)^2}, \\
    \lambda_\ell^{ab} &\equiv \langle a_{\ell m}^a a_{\ell m}^b \rangle  = -\frac{1}{2}\sin 2\theta_\ell C_\ell^0 + \frac{1}{2} \sin 2\theta_\ell C_\ell^{\rm auto} + \cos 2\theta_\ell C_\ell^{\rm cross} = 0.
\end{align}
\end{widetext}

As expected, $\lambda_\ell^{aa}$ and $\lambda_\ell^{bb}$ are the eigenvalues of the signal covariance matrix ${\bf C}_\ell^s$. The frequency-covariance matrix is now given by
\begin{equation}
    {\bf C}_\ell^{\nu\nu} = \mathbf{A}_\ell \mathbf{R}^{-1}_\ell \mathbf{C}_\ell^{r} \mathbf{R}_\ell\mathbf{A}_\ell^t,
\end{equation}
where ${\bf R}_\ell = {\bf R}(\theta_\ell)$ is the rotation matrix defined in Eq.~\eqref{eq:rotation}. The noise contribution to each term is given by 
\begin{equation}
    {\bf n}_\ell = \left[\mathbf{A}^t_\ell \mathbf{R}_\ell \mathbf{N}_\ell^{-1} \mathbf{R}^{-1}_\ell\mathbf{A}_\ell\right]^{-1}.
\end{equation}

In order to assess the detectability of Rayleigh scattering in this framework, we resort to the same formalism as before and introduce a parameter $\alpha$  corresponding to the amplitude of the Rayleigh scattering signal. However, we have to be careful because while the cross-spectrum $\lambda_\ell^{ab}$ is vanishing, its derivative has to be taken while keeping the rotation angle $\theta_\ell$ fixed (even though $\theta_\ell$ does depend on $\alpha$). Hence, we compute 
\begin{equation}
    \frac{\partial {\bf C}_\ell^{\nu\nu}}{\partial \alpha} \Bigr|_{{\bf R}_\ell \: {\rm fixed}} = \mathbf{A}_\ell \mathbf{R}^{-1}_\ell \frac{\partial {\bf C}_\ell^r}{\partial \alpha} \mathbf{R}_\ell\mathbf{A}_\ell^t.
\end{equation}

The derivative of the cross spectrum with respect to our parameter $\alpha$ is then derived to be

\begin{equation}
    \frac{\partial \lambda_\ell^{ab}}{\partial \alpha} \Bigr|_{\substack{\theta_\ell\\ \alpha=1}} = \sin 2\theta_\ell C_\ell^{\rm auto} + \cos 2\theta_\ell C_\ell^{\rm cross} \neq 0.
\end{equation} 

\begin{figure}[!ht]
    \centering
    \includegraphics[width =.48\textwidth]{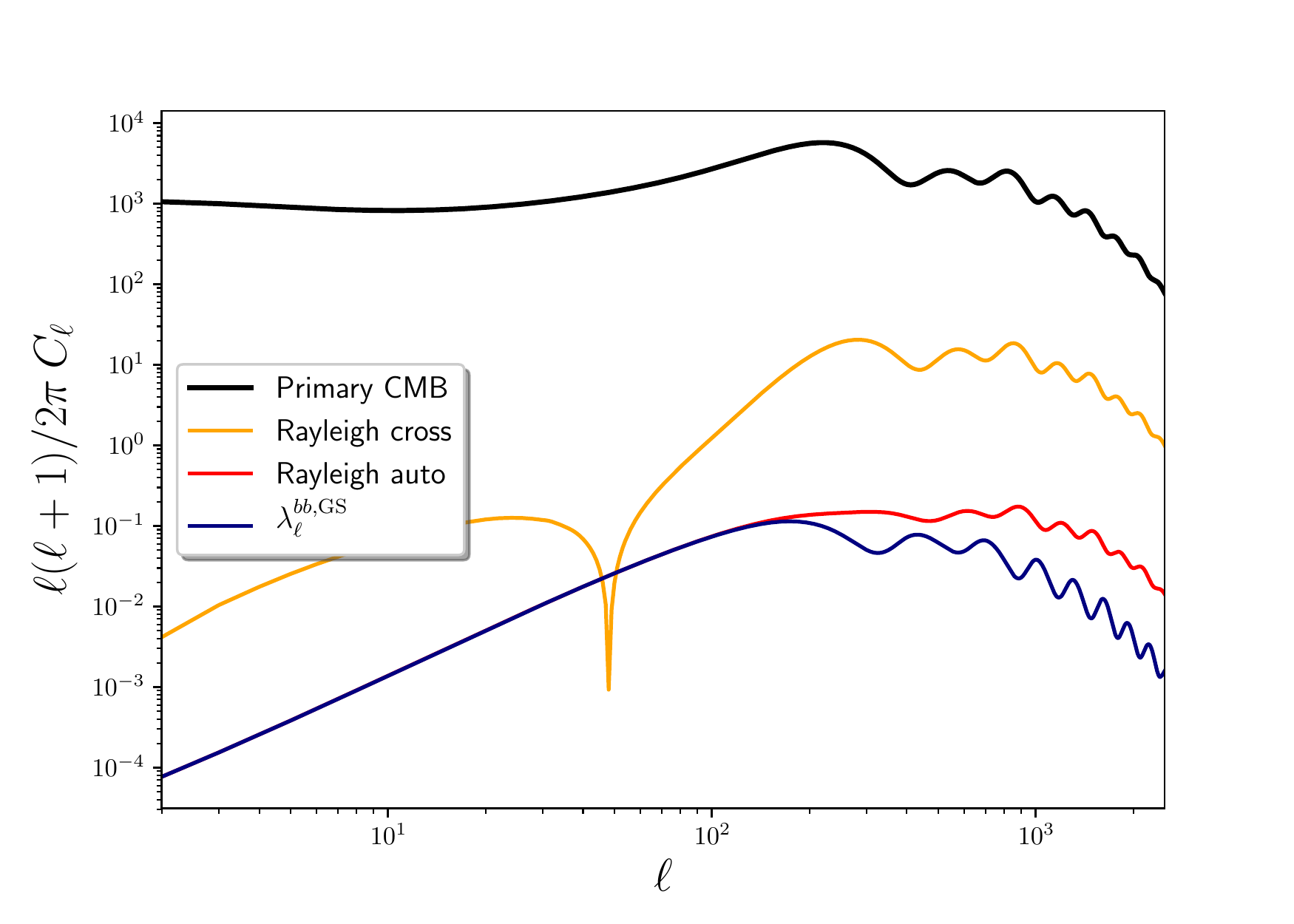}
    \caption{Different temperature spectra discussed in this appendix: Primary CMB (black), Rayleigh scattering cross spectrum (orange), Rayleigh scattering auto spectrum (red), Rayleigh scattering auto spectrum with the primary signal deprojected (blue).}
    \label{fig:deproj_spectra}
\end{figure}

Rotating the basis has the advantage of producing two uncorrelated signals but it does so at the cost of mixing the primary and Rayleigh scattering signals. We could ask whether it would be possible to find another basis, also spanned by uncorrelated signals with one of them remaining unaltered. Such a basis can be derived using the Graham-Schmidt orthogonalization procedure. The same formalism as before can be used by replacing the rotation matrix ${\bf R_\ell}$ with
\begin{equation}
    {\bf R}_\ell^{\rm GS} \equiv 
    \begin{pmatrix}
    1 & 0 \\
    -C_\ell^{\rm cross}/C_\ell^{0} & 1 \\
    \end{pmatrix}.
\end{equation}
The auto- and cross-spectra now read
\begin{align}
    \lambda_\ell^{aa, {\rm GS}} &= C_\ell^0, \\
    \lambda_\ell^{bb, {\rm GS}} &= C_\ell^{\rm auto} - \frac{\left(C_\ell^{\rm cross}\right)^2}{C_\ell^0}, \\
    \lambda_\ell^{ab, {\rm GS}} &= C_\ell^{\rm cross} - C_\ell^0\frac{C_\ell^{\rm cross}}{C_\ell^0} = 0.
\end{align}

As expected, $\lambda_\ell^{aa,GS}$ is the primary CMB while $\lambda_\ell^{bb,GS}$ is the Rayleigh scattering auto spectra deprojected from the correlated part of primary CMB. For the same reason as before, although the cross spectrum is vanishing, its derivative is not 
\begin{equation}
    \frac{\partial \lambda_\ell^{ab,{\rm GS}}}{\partial \alpha} \Bigr|_{\substack{{\rm GS}\\ \alpha=1}} = C_\ell^{\rm cross} \neq 0.
\end{equation} 

 Fig.~\ref{fig:deproj_spectra} shows the temperature spectra defined above. We note that $\lambda_\ell^{bb,GS}$, follows the Rayleigh auto-spectrum on large scale but diverges when the Rayleigh scattering signal is correlated with the primary one (Fig.~\ref{fig:correlation}). We noted earlier that the signal-to-noise of Rayleigh scattering (Eq.~\eqref{eq:SNR_alpha}) is sourced by both the auto- and cross-spectra. After deprojection the signal-to-noise is sourced by both $\lambda_\ell^{bb,GS}$ and our ability to measure $\lambda_\ell^{ab,GS} = 0$. Indeed, if we would apply the deprojection procedure to maps that do not include Rayleigh scattering, we would have $\lambda_\ell^{ab,GS} = - C_\ell^{\rm cross}$. Our ability to detect Rayleigh scattering is therefore equivalent to our ability to reject the null hypothesis above.

\section{Experimental setups}
\label{app:experiments}

\subsection{Noise modelling}
CMB experiments produce noisy measurements of pixels in the sky. The noise is sourced by both the photon noise of the detector arrays and for ground based experiments, by the atmosphere. Assuming a uniform scanning strategy, i.e. each pixel is observed for a constant duration, the noise power spectra can be modelled as~\cite{Ade_2019} 
\begin{equation}
    N_\ell = N_\mathrm{red}\left(\frac{\ell}{\ell_\mathrm{knee}}\right)^{\alpha_\mathrm{knee}} + N_\mathrm{white}.
    \label{eq:noise_model_atmo}
\end{equation}
$N_{\rm white}$ is the {\rm white} noise levels, sourced by photon noise. It is inversely proportional to the number of detectors in the focal plane $N_{\rm det}$ and the integration time $T_{\rm obs}$. When polarization maps are produced, $N_{\rm white}$ is multiplied by two since both the $Q$ and $U$ stokes parameters have to be measured. $N_{\rm red}$ is the {\rm red} noise level, which captures atmospheric noise, and is absent in space-based missions. $\ell_{\rm knee}$ and $\alpha_{\rm knee}$ control how different angular scales are impacted by the atmosphere. For polarization, $N_{\rm red} = N_{\rm white}$.

The amount of noise introduced by the atmosphere is currently mostly based on empirical estimates, which at the SO site are informed by current ACT measurements. At higher altitudes, such as the CCAT-prime site, the atmospheric contamination is not completely determined. It is expected to be smaller since the atmosphere is thinner, but the precise scale dependence and amplitude will only become apparent once the experiment becomes operational.

\subsection{Simons Observatory}
\label{subsec:SO}
The Simons Observatory (SO) is a ground based CMB experiment under construction in the Atacama desert in Chile~\cite{Ade_2019}. It consists of a $6$~m crossed-Dragone large aperture telescope (LAT) targeting small scales anisotropies as well as three $0.5$~m small aperture telescopes (SAT) aiming at detecting primordial gravitational waves. Although SO is located at $5200$~m above sea level, the atmosphere still impacts measurement of both temperature and polarization anisotropies. 

\begin{table}[!ht]
    \centering
    \begin{tabular}{|c|c|c|c|c|}

         \multicolumn{4}{c}{SO SAT ($f_\mathrm{sky} = 0.1$)} \\
         \hline
         Freq. & Beam & $N_\mathrm{white}$ & $\ell_\mathrm{knee}$ & $\alpha_\mathrm{knee}$  \\

         [GHz] & [arcmin] & [$\mu \mathrm{K}^2$] & &  \\
         \hline \hline
         27 & 91 & $5.3\times 10^{-5}$ & 15 & -2.4  \\
         \hline
         39 & 63 & $2.4\times 10^{-5}$ & 15 & -2.49 \\
         \hline
         93 & 30 & $3.0\times 10^{-7}$ & 25 & -2.5 \\
         \hline
         145 & 17 & $3.7\times 10^{-7}$ & 25 & -3.0 \\
         \hline
         225 & 11 & $1.5\times 10^{-6}$ & 35 & -3.0 \\
         \hline
         280 & 9  & $8.5\times 10^{-6}$ & 40 & -3.0 \\
         \hline

    \end{tabular}
    \caption{Characteristics of SO Small Aperture Telescope instruments. SAT will only measure polarization for which $N_\mathrm{red} = N_\mathrm{white}$ and white noise levels should be multiplied by 2 since both Q and U stokes parameters needs to be measured.}
    \label{tab:SO_SAT}
\end{table}

\begin{table}[!ht]
    \centering
    \begin{tabular}{|c|c|c|c|}

         \multicolumn{4}{c}{SO LAT ($f_\mathrm{sky} = 0.4$)} \\
         \hline
         Freq. & Beam & $N_\mathrm{white}$ & $N_\mathrm{red}$ \\
         
         [GHz] & [arcmin] & [$\mu \mathrm{K}^2$] & [$\mu \mathrm{K}^2$] \\
         \hline \hline
         27 & 7.4 & $2.3\times 10^{-4}$ & 100 \\
         \hline
         39 & 5.1 & $6.2\times 10^{-5}$ &  39 \\
         \hline
         93 & 2.2 & $2.8\times 10^{-6}$ &  230\\
         \hline
         145 & 1.4 & $3.6\times 10^{-6}$ &  1500\\
         \hline
         225 & 1.0 & $1.9\times 10^{-5}$ &  17000\\
         \hline
         280 & 0.9 & $1.16\times 10^{-4}$ &  31000\\
         \hline
         
    \end{tabular}
     \caption{Characteristics of SO Large Aperture Telescope instruments. LAT will produce both temperature and polarization maps. In temperature, both $\ell_\mathrm{knee}$ and $\alpha_\mathrm{knee}$ are fixed (resp. 1000 and -3.5). In polarization, $N_\mathrm{red} = N_\mathrm{white}$ and $\ell_\mathrm{knee}=700$ and $\alpha_\mathrm{knee}=-1.4$.}
    \label{tab:SO_LAT}
\end{table}

The use of dichroic detectors also correlates noise spectra from different channels. This results in the noise frequency-covariance matrix no longer being diagonal. This effect has been neglected in our analysis. 

\subsection{CCAT-prime}
\label{subsec:CCAT}

CCAT-prime is a 6~m crossed-Dragone telescope~\cite{CCAT19} targeting high frequency galactic and extra-galactic signals. One of its instruments, Prime-Cam is designed to measure and characterize CMB polarization and foregrounds at high frequencies from a $5600$~m altitude site in the Atacama desert in Chile, close to the SO site. Being a ground-based experiment, CCAT-prime observations will be impacted by the atmosphere (Eq.~\eqref{eq:noise_model_atmo}). Parameters are given in~\cite{Choi19} and are summarized in Table~\ref{tab:CCAT}.

\begin{table}[!h]
    \centering
    \begin{tabular}{|c|c|c|c|}

         \multicolumn{4}{c}{CCAT-prime ($f_\mathrm{sky} = 0.35$)}  \\
         \hline
         Freq. & Beam & $N_\mathrm{white}$ & $N_\mathrm{red}$ \\
         
         [GHz] & [arcsec] & [$\mu \mathrm{K}^2$] & [$\mu \mathrm{K}^2$] \\
         \hline \hline
         220 & 57 & $1.8\times 10^{-5}$ & $1.6\times 10^{-2}$  \\
         \hline
         280 & 45 & $6.4\times 10^{-5}$ & $1.1\times 10^{-1}$  \\
         \hline
         350 & 35 & $9.3\times 10^{-4}$ & $2.7\times 10^{0}$ \\
         \hline
         410 & 30 & $1.2\times 10^{-2}$ & $1.7\times 10^{1}$ \\
         \hline
         850 & 14 & $2.8\times 10^{4}$ & $6.1\times 10^{6}$ \\
         \hline

    \end{tabular}
    \caption{Characteristics of CCAT-prime Prime-Cam instrument. In temperature, both $\ell_\mathrm{knee}$ and $\alpha_\mathrm{knee}$ are fixed (resp. 1000 and -3.5). In polarization, $\ell_\mathrm{knee}$=700 and $\alpha_\mathrm{knee}=-1.4$.}
    \label{tab:CCAT}
\end{table}

\subsection{CMB-S4}
\label{subsec:S4}
CMB-S4 is the Stage-4 ground based CMB experiment with telescopes to be sited in the Chilean Atacama Desert and at the South Pole~\cite{Abazajian:2016yjj,Abazajian:2019eic}.  The current survey design includes a high-resolution wide and deep survey of a large fraction of the sky using two LATs built in Chile, a low-resolution ultra-deep survey conducted by a set of SATs at the South Pole, and a dedicated high-resolution ultra-deep delensing survey conducted by a LAT at the South Pole.  We will focus our attention on the wide survey conducted by the Chile LATs.  We use the survey design matches the current design listed on the public CMB-S4 web page at the time of writing\footnote{\url{https://cmb-s4.org/wiki/index.php/White_noise_levels_for_high_cadence_scan_at_elevation_of_40_degrees}}.

\begin{table}[!ht]
    \centering
    \begin{tabular}{|c|c|c|c|c|}
         \multicolumn{5}{c}{CMB-S4 Chile LATs ($f_\mathrm{sky} = 0.65$)} \\
         \hline
         Freq. & Beam & $N_\mathrm{white}$ & $\ell_\mathrm{knee}$ & $\alpha_\mathrm{knee}$  \\

         [GHz] & [arcmin] & [$\mu \mathrm{K}^2$] & &  \\
         \hline \hline
         27 & 7.4 & $3.9\times 10^{-5}$ & 415 & -3.5  \\
         \hline
         39 & 5.1 & $1.2\times 10^{-5}$ & 391 & -3.5 \\
         \hline
         93 & 2.2 & $3.0\times 10^{-7}$ & 1932 & -3.5 \\
         \hline
         145 & 1.4 & $3.7\times 10^{-7}$ & 3917 & -3.5 \\
         \hline
         225 & 0.9 & $4.0\times 10^{-6}$ & 6740 & -3.5 \\
         \hline
         278 &  0.7 & $2.4\times 10^{-5}$ & 6792 & -3.5 \\
         \hline

    \end{tabular}
    \caption{Characteristics of CMB-S4 instruments. The noise model for CMB-S4 is slightly different than for Simons Observatory and assumes $N_{\rm red} = N_{\rm white}$ for both temperature and polarization. In polarization, $\ell_\mathrm{knee} = 700$, $\alpha_\mathrm{knee} = -1.4$ at all frequencies, and $N_\mathrm{white}$ for polarization is equal to two times that for temperature.}
    \label{tab:S4}
\end{table}

\subsection{LiteBIRD}

LiteBIRD~\cite{Hazumi2019} is a satellite mission that will produce maps of large scale temperature and polarization anisotropies. Primarily targeting a first detection of (primordial) $B$-mode polarization, its dense frequency coverage, combined with its high sensitivity in polarization, will make it a suitable experiment to detect and characterize the Rayleigh scattering signal. Sensitivities and beams are taken from~\cite{Hazumi2019} and are summarised in Table~\ref{tab:LiteBIRD}.

\begin{table}[!ht]
    \centering
    \begin{tabular}{|c|c|c||c|c|c|}
         
          \multicolumn{6}{c}{LiteBIRD ($f_\mathrm{sky} = 0.7$)}   \\
          \hline
         Freq. & Beam & $N_\mathrm{white}$& Freq. & Beam & $N_\mathrm{white}$ \\
        
         [GHz] & [arcmin] & [$\mu \mathrm{K}^2$] & [GHz] & [arcmin] & [$\mu \mathrm{K}^2$]\\
         \hline \hline
         40 & 69 & $2.38\times 10^{-4}$ & 140 & 23 & $5.89\times 10^{-6}$  \\
         \hline
         50 & 56 & $9.75\times 10^{-5}$ & 166 & 21 & $7.15\times 10^{-6}$  \\
         \hline
         60 & 48 & $6.70\times 10^{-5}$ & 195 & 20 & $5.69\times 10^{-6}$  \\
         \hline
         68 & 43 & $4.44\times 10^{-4}$ & 235 & 19 & $1.00\times 10^{-5}$ \\
         \hline
         78 & 39 & $3.08\times 10^{-5}$ & 280 & 24 & $2.95\times 10^{-5}$  \\
         \hline
         89 & 35 & $2.32\times 10^{-5}$ & 337 & 20 & $6.44\times 10^{-5}$  \\
         \hline
         100 & 29 & $1.43\times 10^{-5}$ & 402 & 17 & $2.38\times 10^{-4}$ \\
         \hline
         119 & 25 & $9.77\times 10^{-6}$ &\multicolumn{3}{c}{}   \vline\\
         \hline

    \end{tabular}
    \caption{Characteristics of the LiteBIRD experiment. Noise levels are quoted for temperature maps, polarization noise is a factor two larger at each frequency.}
    \label{tab:LiteBIRD}
\end{table}

\subsection{PICO}
\label{subsec:PICO}
PICO (\emph{Probe of Inflation and Cosmic Origin})~\cite{PICO19} is a proposed satellite mission that will produce high resolution maps of temperature and polarization anisotropies of the CMB. Equipped with 21 frequency channels ($21$-$799$~GHz), it will also contribute to a better understanding of foreground contamination, required for the mitigation of their effects on cosmological observables. These densely packed frequency channels will be a unique opportunity to observe, characterize, and use Rayleigh scattering as a source of cosmological information. Being a spaced-based mission, PICO does not suffer from atmospheric contamination and we take its instrumental noise to be white (convolved with a gaussian beam). Sensitivities and beams are taken from the \emph{Best Current Estimate} in~\cite{PICO19} and are summarised in Table ~\ref{tab:PICO}. 

\begin{table}[!t]
    \centering
    \begin{tabular}{|c|c|c||c|c|c|}
           \multicolumn{6}{c}{PICO ($f_\mathrm{sky} = 0.7$)}  \\
          \hline
         Freq. & Beam & $N_\mathrm{white}$& Freq. & Beam & $N_\mathrm{white}$ \\
         
         [GHz] & [arcmin] & [$\mu \mathrm{K}^2$] & [GHz] & [arcmin] & [$\mu \mathrm{K}^2$]\\
         \hline \hline
         21 & 38.4 & $1.21\times 10^{-5}$ & 159 & 6.2 & $7.15\times 10^{-8}$ \\
         \hline
         25 & 32.0 & $7.15\times 10^{-6}$ & 186 & 4.3 & $3.32\times 10^{-7}$ \\
         \hline
         30 & 28.3 & $3.20\times 10^{-6}$ & 223 & 3.6 & $4.33\times 10^{-7}$ \\
         \hline
         36 & 23.6 & $1.33\times 10^{-6}$ & 268 & 3.2 & $2.05\times 10^{-7}$ \\
         \hline
         43 & 22.2 & $1.33\times 10^{-6}$ & 321 & 2.6 & $3.81\times 10^{-7}$ \\
         \hline
         52 & 18.4 & $6.77\times 10^{-7}$ & 385 & 2.5 & $4.33\times 10^{-7}$ \\
         \hline
         62 & 12.8 & $6.11\times 10^{-7}$ & 462 & 2.1 & $1.73\times 10^{-6}$ \\
         \hline
         75 & 10.7 & $3.81\times 10^{-7}$ & 555 & 1.5 & $4.44\times 10^{-5}$ \\
         \hline
         90 & 9.5 & $1.69\times 10^{-7}$ & 666 & 1.3 & $6.61\times 10^{-4}$ \\
         \hline
         108 & 7.9 & $1.08\times 10^{-7}$ & 799 & 1.1 & $2.32\times 10^{-2}$ \\
         \hline
         129 & 7.4 & $9.52\times 10^{-8}$ &\multicolumn{3}{c}{}   \vline\\
         \hline

    \end{tabular}
    \caption{Characteristics of the proposed PICO experiment. Noise levels are quoted for temperature maps, polarization noise is a factor two larger at each frequency.}
    \label{tab:PICO}
\end{table}

\subsection{DESI}
\label{subsec:DESI}
The information from baryon acoustic oscillations (BAO) is included by adding the information expected from measurements of the the quantity $r_\mathrm{s}/d_\mathrm{V}(z)$, where $r_\mathrm{s}$ is the size of the sound horizon at decoupling, and $d_\mathrm{V}(z)$ is the volume distance to redshift $z$. We use the same procedure as in Ref.~\cite{Allison:2015qca} and calculate Fisher matrix expected from DESI as
\begin{equation}
    F_{ij}^\mathrm{BAO} = \sum_n \frac{1}{\sigma(r_\mathrm{s}/d_\mathrm{V}(z_n))^2}\frac{d(r_\mathrm{s}/d_\mathrm{V}(z_n))}{dp_i}\frac{d(r_\mathrm{s}/d_\mathrm{V}(z_n))}{dp_j} \, .
\end{equation}
The fractional errors on $r_\mathrm{s}/d_\mathrm{V}(z)$ expected from DESI~\cite{Font-Ribera:2013rwa} are given in Table~\ref{tab:DESI}.

\begin{table}[!hbt]
    \centering
    \begin{tabular}{|c|c||c|c|}
         
          \multicolumn{4}{c}{DESI BAO}   \\
          \hline
         Redshift & $\frac{\sigma(r_\mathrm{s}/d_\mathrm{V}(z))}{(r_\mathrm{s}/d_\mathrm{V}(z))}$ [\%] & Redshift & $\frac{\sigma(r_\mathrm{s}/d_\mathrm{V}(z))}{(r_\mathrm{s}/d_\mathrm{V}(z))}$ [\%] \\
         \hline \hline
         0.15 & 1.89 & 1.05 & 0.59 \\ \hline
         0.25 & 1.26 & 1.15 & 0.60 \\ \hline
         0.35 & 0.98 & 1.25 & 0.57 \\ \hline
         0.45 & 0.80 & 1.35 & 0.66 \\ \hline
         0.55 & 0.68 & 1.45 & 0.75 \\ \hline
         0.65 & 0.60 & 1.55 & 0.95 \\ \hline
         0.75 & 0.52 & 1.65 & 1.48 \\ \hline
         0.85 & 0.51 & 1.75 & 2.28 \\ \hline
         0.95 & 0.56 & 1.85 & 3.03 \\ \hline

    \end{tabular}
    \caption{Fractional errors on $r_\mathrm{s}/d_\mathrm{V}(z)$ as a function of redshift expected from the DESI BAO survey.}
    \label{tab:DESI}
\end{table}

All quoted numbers in the main text of this paper are derived using Table~\ref{tab:SO_LAT}-\ref{tab:DESI}. 

\bibliographystyle{apsrev.bst}
\bibliography{Rayleigh}

\end{document}